\documentclass{amsart}

\usepackage{amsmath, color, dsfont}
\usepackage[utf8]{inputenc}
\usepackage{geometry}
\usepackage{paralist} 
\usepackage{subfig} 
\usepackage[usenames,dvipsnames]{xcolor}
\usepackage{graphicx}
\usepackage{feynmf}
\usepackage[all]{xy}

\usepackage{amsbsy,amssymb,amscd,amsfonts,latexsym,amstext,delarray,
amsmath,graphicx} 
\input xypic
\usepackage{color}

\newtheorem{thm}{Theorem}[section]
\newtheorem{prop}[thm]{Proposition}

\newtheorem{lem}[thm]{Lemma}

\newtheorem{defn}[thm]{Definition}

\newtheorem{ex}[thm]{Example}
\newtheorem{ques}[thm]{Question}

\numberwithin{equation}{section}

\def\C{{\mathbb C}}

\def\N{{\mathbb N}}

\def\Z{{\mathbb Z}}
\def\R{{\mathbb R}}
\def\K{{\mathbb K}}

\def\cA{{\mathcal A}}
\def\cB{{\mathcal B}}

\def\cD{{\mathcal D}}

\def\cH{{\mathcal H}}
\def\cI{{\mathcal I}}

\def\cM{{\mathcal M}}

\def\cO{{\mathcal O}}
\def\cP{{\mathcal P}}

\def\cR{{\mathcal R}}

\def\cV{{\mathcal V}}

\def\c{{\mathfrak{c}}}
\def\b{{\mathfrak{b}}}
\def\r{{\mathfrak{r}}}

\title{Dyson--Schwinger equations in the theory of computation}
\author{Colleen Delaney and Matilde Marcolli}
\address{Physics Department, Caltech, 1200 E.~California Blvd. Pasadena, CA 91125, USA}
\email{cdelaney@caltech.edu}
\address{Mathematics Department, Caltech, 1200 E.~California Blvd. Pasadena, CA 91125, USA}
\email{matilde@caltech.edu} 

\begin{document}

\begin{abstract}
Following Manin's approach to renormalization in the theory of computation, we
investigate Dyson--Schwinger equations on Hopf algebras, operads and properads
of flow charts, as a way of encoding self-similarity structures in the theory of algorithms
computing primitive and partial recursive functions and in the halting problem. 
\end{abstract}

\maketitle

\section{Introduction}

There are many deep connections between theoretical physics and information theory,
and in particular the theory of computation, see for instance the account given in \cite{Baez}.
In the recent papers \cite{Man1} and \cite{Man2}, Manin developed a new approach to
the theory of computation and the halting problem, based on importing ideas and techniques
from the Hopf-algebraic formulation of renormalization in perturbative quantum field theory,
\cite{CK}, \cite{EFGK}, \cite{Kr}. The purpose of this paper is to show that the Hopf
algebra of flow charts introduced by Manin in \cite{Man1} exhibits self-similarity structures
given by solutions of combinatorial Dyson--Schwinger equations, defined as in 
perturbative renormalization, \cite{BergKr}, \cite{Foissy1}, \cite{Foissy2}, 
\cite{Foissy3}, \cite{KrYeats}, \cite{Yeats}.  These can be thought of as a notion 
of ``equations of motion" in the theory of computation.

\smallskip

Dyson--Schwinger equations \cite{Dyson}, \cite{Schwinger}
are a formulation of equations of motion in perturbative
quantum field theory, expressed in the form of relations between Green functions. 
In recent years, an algebraic formulation of the combinatorial structure of perturbative
renormalization for scalar field theories was developed, starting with the work of
Kreimer on the Hopf-algebraic structure of renormalization \cite{Kr}, followed by the
Connes--Kreimer formulation \cite{CK} of the BPHZ renormalization procedure,
and the formulation in terms of a Riemann--Hilbert correspondence for categories of 
differential systems in Connes--Marcolli \cite{CoMa}, \cite{CoMa2}. The Connes--Kreimer
formulation of renormalization was given a very general algebraic formulation in terms of
Rota--Baxter algebras in the work of Ebrahimi-Fard, Guo and Kreimer \cite{EFGK}. 
Correspondingly, the Dyson--Schwinger equations were also given a combinatorial
form, reflecting the Hopf-algebraic structure of renormalization, in the work of
Bergbauer, Kreimer, and Yeats,  \cite{BergKr}, \cite{KrYeats}, \cite{Yeats} and 
more recently with an extensive study by Foissy, \cite{Foissy1}, \cite{Foissy2}, \cite{Foissy3}.

\smallskip

In this paper we investigate the formulation of Dyson--Schwinger equations
in the context of Manin's approach to renormalization and computation.
Our main application will be an alternative formulation of a possible ``Galois
theory of algorithms", similar in spirit to the approach proposed by Yanofsky
in \cite{Yano2}, but more directly related to the Hopf algebra structure of
algorithms.

\smallskip

In \S \ref{PrimRecSec} we review some variants of the construction of a Hopf
algebra of {\em flow charts}. These are diagrams, consisting of decorated planar
rooted trees, that compute primitive recursive functions. 
In \S \ref{algosec}, we discuss how different version of the Hopf algebra, of
flow charts or of ``templates", give rise to slightly different notions of {\em algorithms}, 
meant in the sense of \cite{Yano} as intermediate levels between programs and 
functions. 
In \S \ref{DSflowSec} we
discuss grafting operators and Dyson--Schwinger equations for flow charts,
as a way of identifying certain self-similarity structures in the computation of
primitive recursive functions. We review some known results about existence
and uniqueness of solutions and on conditions under which the coordinates
of the solution generate a Hopf subalgebra. We show that one can also consider
ideals generated by the coordinates of solutions. Under the condition that 
the grafting operator is a cocycle, these can give rise to Hopf ideals. 
The quotient Hopf algebra determines a subgroup of the affine group scheme dual to the
(commutative) Hopf algebra of flow charts. %
We propose this subgroup as a suitable
notion of Galois group, following a point of view closely related to 
Yanofsky's Galois theory of algorithm \cite{Yano2}, although the groups we
consider are of a different nature than those considered in \cite{Yano2}. The
group we consider consists %
of those symmetries that are compatible with the self-similarity structure imposed
by the Dyson--Schwinger equation. We then show an explicit example
of a Dyson--Schwinger equation based on computation by binary trees that
does not satisfy the Hopf algebra condition. In order to accommodate this
type of example, that is relevant to the setting of computation, we show
that it is convenient to take an operadic viewpoint on Dyson--Schwinger equations,
which was already suggested in the work of Bergbauger--Kreimer \cite{BergKr}.
In \S \ref{operadSec} we show that Dyson--Schwinger equations can be 
formulated in the operad setting, using a family of grafting operations, which
is not required to satisfy a cocycle condition. We consider 
Dyson--Schwinger equations on an {\em operad of flow charts}. We prove the
existence and uniqueness of solutions for these operadic Dyson--Schwinger 
equations and also for a {\em properad} version. The extension from operads
to properads is motivated by the more recent formulation, given by Manin in
\cite{Man-Zipf}, of a {\em properad of flow charts} based on directed (acyclic) graphs
instead of rooted trees. 
Finally, \S \ref{haltSec} is more speculative in nature and it focuses on an algebraic
Feynman rule associated to the halting problem proposed in \cite{Man2}.
We also formulate some general questions about the meaning of the 
BPHZ renormalization procedure and the role of Dyson--Schwinger equations in the
halting problem.

\section{Primitive recursive functions and the Hopf algebra of flow charts}\label{PrimRecSec}

We recall here the setup of Manin \cite{Man1} on the Hopf algebra of flow charts.
First recall from \cite{Man-logic} the following facts about primitive recursive functions.

\medskip

\subsection{Primitive recursive functions}

As in \S V.2 of \cite{Man-logic}, we consider the class of {\em primitive recursive functions}
as generated by the {\em basic functions}
\begin{itemize}
\item Successor $s: \N \to \N$, $s(x)=x+1$;
\item Constant $c^n: \N^n \to \N$, $c^n (x)=1$ (for $n\geq 0$);
\item Projection $\pi_i^n: \N^n \to \N$, $\pi_i^n(x)=x_i$ (for $n\geq 1$);
\end{itemize}
With the {\em elementary operations}
\begin{itemize}
\item Composition (substitution) $\c_{(m,m,p)}$: for $f: \N^m \to \N^n$, $g: \N^n \to \N^p$,
$$ g\circ f : \N^m \to \N^p, \ \ \  \cD(g\circ f)=f^{-1}(\cD(g)); $$
\item Bracketing (juxtaposition) $\b_{(k,m,n_i)}$: for $f_i: \N^m \to \N^{n_i}$, $i=1,\ldots,k$,
$$ f=(f_1,\ldots, f_k): \N^m \to \N^{n_1+\cdots+n_k}, \ \ \  \cD(f)=\cD(f_1)\cap\cdots\cap \cD(f_k); $$
\item Recursion $\r_n$: for $f: \N^n \to \N$ and $g: \N^{n+2} \to \N$,
$$ h(x_1,\ldots, x_n,1):= f(x_1,\ldots, x_n), $$
$$ h(x_1,\ldots, x_n,k+1):=g(x_1,\ldots,x_n,k,h(x_1,\ldots, x_n,k)), \ \ \ k\geq 1, $$
where recursively $(x_1,\ldots, x_n,1)\in \cD(h)$ iff $(x_1,\ldots, x_n)\in \cD(f)$ and
$(x_1,\ldots, x_n,k+1) \in \cD(h)$ iff $(x_1,\ldots,x_n,k,h(x_1,\ldots, x_n,k)\in \cD(g)$.
\end{itemize}

\medskip

\subsection{Hopf algebras of decorated rooted trees}

We recall here the construction of Hopf algebras of decorated rooted trees and
of planar decorated rooted trees and their main properties, see \cite{Foissy2}
and references therein. Throughout this paper vector spaces and algebras are
over a field $\K$ of characteristic zero.

\subsubsection{Rooted trees} A rooted tree is a finite graph $\tau$, whose geometric realization $|\tau|$ 
is simply connected, defined by combinatorial
data $(F_\tau, V_\tau, v_\tau, \delta_\tau, j_\tau)$, with $F_\tau$ a set of
half-edges (also called flags or tails), $V_\tau$ the set of vertices with a
distinguished element $v_\tau\in V_\tau$ (the root), a boundary map 
$\partial_\tau: F_\tau \to V_\tau$ that associates to half-edge its boundary vertex 
and an involution $j_\tau: F_\tau \to F_\tau$, $j_\tau^2=1$ that 
performs the matching of half-edges that determines the edges of $\tau$. 
The resulting graph has edges (or internal edges) given by
pairs of half-edges matched by the involution, and tails (or external edges)
given by half-edges that are fixed by the involution. We denote by $E_\tau$
the set of internal edges and by $E^{ext}_\tau$ the set of external edges.

\subsubsection{Orientations} We consider rooted trees as endowed with the orientation that the root 
vertex as the {\em output}, namely where all edges are oriented in the
direction of the unique path to the root. 

\subsubsection{External edges} 
We also assume that the root vertex has an outgoing half-edge and a number
of incoming edges, while all other vertices have one outgoing
edge and a number of incoming edges and possibly a number of 
incoming tails. Each leaf of the tree has an incoming tail.

\subsubsection{Planarity}
A {\em planar} rooted tree is a rooted tree $\tau$ together with a fixed embedding 
$\iota: |\tau|\hookrightarrow \R^2$ of its geometric realization in the plane.

\subsubsection{Decorations}
A (planar) decorated rooted tree is rooted tree and with a map $\phi_V: V_\tau \to \cD_V$ of the set
of vertices to a set $\cD_V$ of vertex-labels and a map $\phi_F: F_\tau \to \cD_F$ to a set
$\cD_F$ of labels of flags. The assignment of a decoration $(\phi_V,\phi_F)$ to a rooted tree $\tau$ is in general subject to constraints. For example, if $f$ and $f'$ in $F_\tau$ are matched by the
involution $j_\tau$ to form an edge $e\in E_\tau$, then they should carry the same decoration
$\phi_F(f)=\phi_F(f')$. This defines a labeling of edges. We will see other constraints below,
in the specific case of the Hopf algebra of flow charts. We also consider the forgetful map
that forgets the flag labels $\cD_F$ and retains the vertex labels $\cD_V$ and a further
forgetful map to unlabeled trees.

\subsubsection{Noncommutative Hopf algebra of planar rooted trees}

Given a (planar) rooted tree $\tau$, an {\em admissible cut} $C$ of $\tau$ is a
modification of the involution $j_\tau$ that {\em cuts} a subset of internal edges 
$e_i \in E_\tau$ into two flags $f_i$, $f_i'$. Namely, instead of having $j_\tau(f_i)=f_i'$,
one modifies the involution so that the new $\hat j(f_i)=f_i$ and $\hat j(f_i)=f_i'$,
where the set of edges $e_i$ is chosen in such a way that every oriented path in
$\tau$ from a leaf vertex to the root contains at most one $e_i$. The new graph
obtained using the involution $\hat j$ is a forest 
$$ C(\tau) = \rho_C(\tau) \amalg \pi_C(\tau) $$
consisting of one (planar)
rooted tree $\rho_C(\tau)$ containing the root vertex of $\tau$ and a finite disjoint union of other
oriented planar trees $\pi_C(\tau)= \amalg_i \pi_{C,i}(\tau)$, where each tree $\pi_{C,i}(\tau)$ has 
a single output external edge, 
to which we assign as root the boundary vertex of this output.

\smallskip

The {\em noncommutative} Hopf algebra of planar rooted trees $\cH^{nc}$ is defined, as an algebra,
as the free algebra generated by the planar rooted trees, with the 
coalgebra structure given by the {\em admissible-cuts} coproduct
\begin{equation}\label{cutcoprod}
\Delta(\tau) = \tau \otimes 1 + 1 \otimes \tau +\sum_C \pi_C(\tau) \otimes \rho_C(\tau).
\end{equation}
The coproduct is coassociative and the Hopf algebra is graded by the number of vertices 
of rooted trees, hence the antipode is defined inductively by the formula
$$ S(x) = -x - \sum S(x') x'', \ \ \text{ for } \Delta(x) = x\otimes1 +1 \otimes x + \sum x' \otimes x'' $$
with $x'$, $x''$ terms of lower degree. See \cite{Foissy4}, \cite{Holt} for more details on this
Hopf algebra.
An element $x$ in a Hopf algebra is primitive if 
$\Delta(x) = 1\otimes x + x \otimes 1$.

\subsubsection{Commutative Hopf algebra of rooted trees}

In the Hopf algebraic approach to perturbative renormalization, as formulated
by Kreimer \cite{Kr} and Connes--Kreimer \cite{CK}, one considers a {\em commutative}
Hopf algebra of rooted trees decorated by Feynman graphs, with the admissible-cuts
coproduct \eqref{cutcoprod}, or else a commutative Hopf algebra of Feynman graphs, with 
a coproduct that corresponds to inclusions of subgraphs (subdivergences) and quotient graphs.

\smallskip

In the case of planar rooted trees, one can also consider a commutative Hopf
algebra $\cH$, which is a quotient of $\cH^{nc}$ by imposing the commutativity
of multiplication. The generators of $\cH$ are still planar rooted trees, but monomials
in these generators no longer identify with {\em embedded} forests. The grading and
antipode are as above, induced by those of $\cH^{nc}$.

\medskip

\subsection{Manin's Hopf algebra of flow charts}

We consider, as in \cite{Man1}, the set of planar labelled rooted trees,
where the label set $\cD_V$ of vertices is given by the set of elementary operations 
$$\{ \c_{(m,n,p)}, \b_{(k,m,n_i)}, \r_n \}$$ 
(composition, bracketing, recursion) and the label set $\cD_F$ of the flags is 
the set of primitive recursive functions.
Notice that, because the inputs of the composition,
bracketing and recursion operations are ordered, 
we need to work with {\em planar} rooted trees. 

\smallskip

The data $(\phi_V(v), \phi_F(f_i))$ of a labelings  
$\phi_V(v)\in \cD_V$ and $\phi_F(f_i)$ of a vertex $v\in V_\tau$ and of all 
the half-edges $f_i\in F_\tau$ with $\partial(f_i)=v$ are 
{\em admissible} if they satisfy the following conditions:
\begin{itemize}
\item If $\phi_V(v)=\c_{(m,n,p)}$, then $v$ must have valence three; the labels $h_1=\phi_F(f_1)$
and $h_2=\phi_F(f_2)$ of the two incoming flags must have domains and ranges 
$h_1: \N^m \to \N^n$ and $h_2: \N^n\to \N^p$ and the outgoing flag must be labeled by
the composition $h_2\circ h_1=\c_{(m,n,p)}(h_1,h_2)$.
\item If $\phi_V(v)=\r_n$, then $v$ must have valence three; the labels $h_1=\phi_F(f_1)$
and $h_2=\phi_F(f_2)$ of the two incoming flags must have domains and ranges 
$h_1: \N^n \to \N$ and $h_2: \N^{n+2} \to \N$ and the outgoing flag must be labeled by the
recursion $h=\r_n(h_1,h_2)$.
\item If $\phi_V(v)=\b_{(k,m,n_i)}$ then $v$ must have valence $k+1$ and all the functions 
$h_i=\phi_F(f_i)$ associated to the incoming flags must have domain in the same $\N^m$.
The outgoing flag must be labeled by the bracketing $f=(f_1,\ldots,f_k)=\b_{(k,m,n_i)}(f_1,\ldots,f_k)$.
\end{itemize}

\begin{defn}\label{admlabel}
An {\em admissible labeling} of a planar rooted tree $\tau$ by primitive recursive functions is
a collection of labelings 
$$ \{ (\phi_V(v), \phi_F(f_i))\,|\, v\in V_\tau \, \text{  and } \, f_i\in F_\tau \, \text{ with } \,
\partial(f_i)=v \} $$
such that the admissibility condition above is satisfied at all vertices and such that
$\phi_F(f)=\phi_F(f')$ whenever $j_\tau(f)=f'$.

\noindent A {\em vertex labeling} of a planar rooted tree $\tau$ by elementary operations
is a collection of labeling $\{ \phi_V(v) \in \{ \c, \r, \b_k \} \, | \, v\in V_\tau \}$ with the only constraint
that labels of type $\c$ and $\r$ can only be assigned to vertices with two incoming
flags and one outgoing flag and labels of type  $\b_k$ can be assigned to vertices
with $k$ incoming flags and one outgoing flag.
\end{defn}

Via the forgetful map that keeps the vertex labels and forgets the flag labels, 
a planar rooted tree with an admissible labeling by primitive recursive functions determines
one with vertex labeling by elementary operations, while not all vertex labeling by 
elementary operations on a given planar rooted tree will admit a compatible flag
labeling. Notice that, in the case of vertex labelings, we no longer distinguish between
labels $\c_{(m,n,p)}$, with different choices of $m$, $n$ and $p$, since we do not
fix the flag labels: all these vertex labels in $\cD_V$ correspond to the same label $\c$.
Similarly, all the labels $\r_n$ of admissible labeling correspond to the same label $\r$
of vertex labeling, and all the $\b_{(k,m,n_i)}$ correspond to the same $\b_k$. We retain
the different $\b_k$ as vertex labelings as those correspond to different valences of the vertex.

\smallskip

Thus, a planar rooted tree with an admissible labeling by primitive recursive functions
can be interpreted as a {\em flow chart} that computes the output function starting from the
functions associated to the incoming external edges according to the operations
performed at the vertices.

A planar rooted tree $\tau$ with a vertex labeling by elementary operations, on the other
hand, should be thought of as a template of a possible computational scheme:
when applied to a tuple of primitive recursive functions $(h_1, \ldots, h_k)$, with
$k$ the number of incoming flags of $\tau$, it either computes an output functions $h$,
if the inputs $(h_1, \ldots, h_k)$ together with the assigned vertex labels determine
an admissible labeling of $\tau$ by partial recursive function, or else we set its output
to be the empty function. 

\medskip

We then define two slightly different versions of the Hopf algebra of flow
charts, depending on the type of labeling that we want to use on trees.

\smallskip

\begin{defn}\label{flowHopf}
The {\em noncommutative} Hopf algebra $\cH^{nc}_{{\rm flow},\cP}$
of flow charts is the free algebra generated by the
planar rooted trees with admissible labelings by primitive recursive functions, with the
coproduct \eqref{cutcoprod}. 
We also denote by $\cH_{{\rm flow},\cP}$ the commutative quotient of $\cH^{nc}_{{\rm flow},\cP}$.
Similarly, the noncommutative Hopf algebra $\cH^{nc}_{{\rm flow},\cV}$ is the free algebra
generated by planar rooted trees with vertex labeling by elementary operations, again
with the admissible-cuts coproduct. We denote by $\cH_{{\rm flow},\cV}$ its commutative
quotient. 
\end{defn}

Notice that if $\tau$ has an admissible labeling, then the
rooted trees $\rho_C(\tau)$ and $\pi_{C,i}(\tau)$ also inherit induced admissible labelings,
so the Hopf algebras are well defined.  There is a Hopf algebra homomorphism
$\cH^{nc}_{{\rm flow},\cP} \to \cH_{{\rm flow},\cV}$, determined 
by forgetting the flag labels, but as observed above not all generators of $\cH_{{\rm flow},\cV}$ are
in the image of this map.  Since the rooted trees in $\cH_{{\rm flow},\cV}$ have no labels
attached to flags, we can represent them as rooted trees without external edges. This will
be implicitly done in the description of the grafting operators in Lemma \ref{1cocycle1} below.

\smallskip

In the following, whenever a statement applies to both types of tree decorations on the
Hopf algebras of flow charts, we will drop the $\cP$ and $\cV$ subscripts.

\medskip

\subsubsection{Binary versus $k$-ary operations}

It is possible to reformulate the Hopf algebra of flow charts by using only {\em binary}
trees. In fact, it is shown in \S 2 of \cite{Yano} that the description of 
primitive recursive functions in terms of  basic functions (successor, constants and
projections) and elementary operations (composition, bracketing and recursion),
as recalled above, can be reformulated as repeated applications of {\em binary} operations
of these same forms. In fact, the composition and recursion operations already
are binary (they label vertices with two input flags and one output) so one only
needs to verify that the bracketing operation $\b_{(k,m,n_i)}$ assigning to $k$
functions $f_i: \N^m \to \N^{n_i}$, $i=1,\ldots,k$ their juxtaposition 
$f=(f_1,\ldots, f_k): \N^m \to \N^{n_1+\cdots+n_k}$ is a composition
$$ \b_{(k,m,n_i)} = \b_{(2,m,n_1,n_2+\cdots+n_k)} \circ \cdots \circ \b_{(2,m,n_{k-1},n_k)}. $$
This replaces a single vertex with $k$ inputs and one output with $k$ vertices with two inputs
and one output.
Thus, without loss of generality in the algorithmic representation of primitive recursive functions,
one can consider a sub-Hopf algebra $\cH^{nc,0}_{\rm flow}\subset \cH^{nc}_{\rm flow}$
generated by {\em binary} planar rooted tree with an admissible labeling by primitive 
recursive functions, and the corresponding commutative quotient $\cH^0_{\rm flow}\subset 
\cH_{\rm flow}$. 

\medskip

One the other hand, one can also extend the two binary operations of composition
and recursion to $k$-ary operations for arbitrary $k\geq 2$. In the case of composition,
we can introduce a new set of vertex labels $\c_{(k,n_i)}$ that correspond to the
$k$-ary compositions $\c_{(k,m,n_i)}(h_i)= h_k\circ \cdots \circ h_1$ of
functions $h_i: \N^{n_{i-1}} \to \N^{n_i}$, for $i=1,\ldots, k$, with $n_0=m$. 
In the case of recursion, for $k\geq 1$, one can consider $(k+1)$-ary recursions where the
recursion depends on $k$ initial conditions. Namely, we define $\r_{k+1,n}$
as the operation that computes the function $h=\r_{k,n}(h_1,\ldots, h_{k+1}): \N^{n+1}\to \N$,
from the input $h_i:\N^n \to \N$ for $i=1,\ldots,k$ and $h_{k+1}:\N^{2n+1}\to \N$, as 
\begin{equation}\label{karyrec}
\begin{array}{l}
h(x_1,\ldots,x_n,1)=h_1(x_1,\ldots,x_n), \\
\vdots \\
h(x_1,\ldots,x_n,k)=h_k(x_1,\ldots,x_n), \\
h(x_1,\ldots,x_n,k+\ell)=h_{k+1}(x_1,\ldots,x_n,h_1(x_1,\ldots,x_n),\ldots,h_k(x_1,\ldots,x_n),k+\ell-1), \\
\text{for } \ell \geq 1
\end{array}
\end{equation}

These $k$-ary operations are clearly reducible to compositions of binary
operations, hence any primitive recursive function that can be computed using
these operations can also be computed by the presentation in terms of
binary compositions and recursions. However, as we shall discuss more in
detail below, it is convenient to include these explicitly in the structure of
the Hopf algebra of flow charts in order to have well defined grafting operators
that give rise to Hochschild cocycles and to Dyson--Schwinger equations.

\smallskip

Thus, we propose another minor modification of Manin's Hopf algebra
of flow charts, as follows. 

\begin{defn}\label{flowHopf2}
We denote by $\cH^{nc}_{{\rm flow},\cP'}$ and $\cH^{nc}_{{\rm flow},\cV'}$
the noncommutative Hopf algebras of flow charts where the vertex label sets is
enlarged to contain all the $k$-ary compositions and recursions $\c_{(k,n_i)}$ and 
$\r_{k,n}$, and the admissibility condition is restated accordingly. We denote
by $\cH_{{\rm flow},\cP'}$ and $\cH_{{\rm flow},\cV'}$ their commutative quotients.
\end{defn}

\medskip

\section{Flow charts, templates, and algorithms}\label{algosec}

In the above, we have introduced different versions of rooted trees with 
vertex and edge decorations. We discuss here briefly the meaning of
the objects considered above from the point of view of computation.

\smallskip
\subsection{Flow chartes versus templates}

A rooted tree with vertices 
decorated by elementary operations and edges compatibly
decorated by recursive functions is a {\em flow chart}. It 
describes a concrete realization of a primitive recursive function (the output)
in terms of a computation that starts with other primitive recursive functions,
possibly basic (the inputs of the tree) with the operations performed at
the vertices.

\smallskip

By extension, we also refer to arbitrary elements of the Hopf algebra
$\cH^{nc}_{{\rm flow},\cP}$ as ``flow charts". These are formal sums of forests, with
a planar embedding, decorated as above by primitive recursive functions
at edges and elementary operations at vertices. In this extended sense
a flow charts computes a formal sum of ordered sequences of primitive recursive
functions, given by the collection of the outputs of each of the trees.

\smallskip

We also considered a different type of objects, given by rooted trees
where only the vertices are decorated by elementary operations,
while the edges are undecorated. These are computational
architectures where the inputs are open: once a set of primitive
recursive functions is assigned as input of such a tree, one obtains 
a primitive recursive function as output, computed according to
the operations specified at the vertices. We refer to these
trees with vertex decorations as {\em templates}.

\smallskip

By extension we also refer to an arbitrary element of the Hopf algebra
$\cH^{nc}_{{\rm flow},\cV}$ as a {\em template}. These are formal sums
of forests with vertex decorations by elementary operations. 

\smallskip

The difference between working with flow charts (vertex and edge decorations)
or with what we call here templates (vertex decorations only) will become more
transparent when one considers implementing certain ``reasonable equivalence
relations", which describe, in the general philosophy of Yanofsky \cite{Yano},
an intermediate level of {\em algorithms}, in between programs (flow charts) and
primitive recursive functions.

\smallskip
\subsection{The notion of algorithm}

Yanofsky's work \cite{Yano} provides an in depth discussion of the
notion of {\em algorithm}. To summarize quickly his general approach,
one considers equivalence classes consisting of different programs that
implement ``the same algorithm". In this view, algorithms describe any
intermediate level between the programs (the flow charts in our language
above) and the primitive recursive functions they compute. Rather than
fixing precisely what the intermediate level should be, Yanofsky
allows for some freedom, by considering all the possible quotients
of the set of programs by ``reasonable equivalence relations" and
organizing all these intermediate levels into a hierarchy governed
by a suitable kind of Galois correspondence, \cite{Yano}, \cite{Yano2}.

\smallskip

In Section 3 of \cite{Yano}, Yanofsky gives a list of examples of
``reasonable equivalence relations" one can impose on 
programs. These include: associativity of composition, projections
acting as identity, distributivity of composition over bracketing, associativity
of bracketing, almost-commutativity of bracketing, functoriality of bracketing,
idempotency of twists, Reidemeister moves, relation between recursion and
bracketing, relation between recursion and composition, relation between
recursion and the successor function. 

\smallskip

Our viewpoint here follows closely this idea of Yanofsky, but with
one important difference. We consider the existence of a Hopf
algebra structure on flow charts as an essential part of the data
and we therefore require that a notion of ``reasonable equivalence
relation" should include the compatibility with the Hopf algebra
structure. This means requiring that equivalence relations among
flow charts should define Hopf ideals.

\smallskip

Notice how some of the relations listed above from \cite{Yano} 
reflect properties (associativity, distributivity, almost-commutativity) of the
elementary operations, which are independent of the input
functions, while others involve relations between the elementary
operations and certain specific input functions (projections, 
successor). The different nature of these two types of relations
will be reflected, in our viewpoint, in the fact that the first type of 
relations can be implemented both at the level of the Hopf
algebra of flow charts $\cH^{nc}_{{\rm flow},\cP}$ and at the
level of the Hopf algebra of templates $\cH^{nc}_{{\rm flow},\cV}$,
while the second type of relation can only be implemented 
in $\cH^{nc}_{{\rm flow},\cP}$.

\smallskip

To see a concrete example of both cases, let us first consider
the relation that corresponds to associativity of composition, as
stated in \S 3.1.1 of \cite{Yano}. This implies identifying 
trees of the form
\begin{equation}\label{assrel}
\tau_1= \xy
\POS(6,14) \ar^{g}@{-}+(-4,-8)
\POS(-2,14) \ar_{f}@{-} +(4,-8)
\POS(2,6) *+{\bullet}*+{   \hskip 15pt  \textbf{c}}
\POS(2,6) \ar_{f\circ g}@{-} +(0,-8)
\POS(2,-2) \ar_{}@{-} +(4,-8)
\POS(10,-2) \ar^{h}@{-}+(-4,-8)
\POS(6,-10) *+{\bullet}*+{   \hskip 15pt  \textbf{c}}
\POS(6,-10) \ar_{f\circ g \circ h}@{-}+(0,-8)
\endxy
\ \ \ \ \ \ \cong \ \ \ \ \ \ 
\xy
\POS(6,14) \ar^{h}@{-}+(-4,-8)
\POS(-2,14) \ar_{g}@{-} +(4,-8)
\POS(2,6) *+{\bullet}*+{   \hskip 15pt  \textbf{c}}
\POS(2,6) \ar_{g\circ h}@{-} +(0,-8)
\POS(2,-2) \ar_{}@{-} +(-4,-8)
\POS(-2,-10) \ar^{f}@{-}+(-4,8)
\POS(-2,-10) *+{\bullet}*+{   \hskip 15pt  \textbf{c}}
\POS(-2,-10) \ar_{f\circ g \circ h}@{-}+(0,-8)
\endxy = \tau_2
\end{equation}

In order to implement this relation at the level of the
Hopf algebra, we need to construct a Hopf ideal.
The ideal $\cI=(\tau_1-\tau_2)$ in the underlying algebra 
would not, by itself, be a Hopf ideal, because the 
non-primitive part of the coproduct $\Delta(\tau_1 -\tau_2)$ 
would not be in $\cH \otimes \cI \oplus \cI\otimes \cH$,
but if we also consider the relation identifying the
trees 
\begin{equation}\label{assrel2}
\tau_1' = \xy
\POS(2,6) \ar_{f}@{-} +(0,-8)
\POS(2,-2) *+{\bullet}*+{   \hskip 15pt  \textbf{c}}
\POS(2,-2) \ar_{f}@{-} +(4,-8)
\POS(10,-2) \ar^{g}@{-}+(-4,-8)
\POS(6,-10) *+{\bullet}*+{   \hskip 15pt  \textbf{c}}
\POS(6,-10) \ar_{f\circ g}@{-}+(0,-8)
\endxy
\ \ \ \ \ \ \cong \ \ \ \ \ \
\xy
\POS(2,6) \ar_{g}@{-} +(0,-8)
\POS(2,-2) *+{\bullet}*+{   \hskip 15pt  \textbf{c}}
\POS(2,-2) \ar_{g}@{-} +(-4,-8)
\POS(-2,-10) \ar^{f}@{-}+(-4,8)
\POS(-2,-10) *+{\bullet}*+{   \hskip 15pt  \textbf{c}}
\POS(-2,-10) \ar_{f\circ g}@{-}+(0,-8)
\endxy = \tau_2'
\end{equation}
and the ideal $\cI=(\tau_1-\tau_2, \tau'_1-\tau'_2)$, then
the coproduct satisfies $\Delta(\cI)\subset \cH \otimes \cI 
\oplus \cI\otimes \cH$. The resulting Hopf ideal is the
appropriate way, in our setting, to implement the ``associativity
of composition" relation.

\smallskip

Notice that the relations defining $\cI$ do not depend
on particular input functions being assigned to the
incoming edges, hence they not only define a Hopf ideal 
in the Hopf algebra $\cH^{nc}_{{\rm flow},\cP}$ of flow
charts, but they also define a Hopf ideal in the Hopf
algebra $\cH^{nc}_{{\rm flow},\cV}$ of templates.

\smallskip

We then consider, as a second example, the relation
defined in \S 3.1.2 of \cite{Yano}, which represents
the fact that projection $\pi^n=( \pi^n_i ): \N^n \to \N^n$
is the identity. This relation is implemented by identifying,
for any primitive recursive function $f: \N^n \to \N^m$,
the decorated rooted trees
\begin{equation}\label{relproj}
\xy
\POS(6,14)\ar^{\hskip 8pt f}@{-}+(-4,-8) 
\POS(-2,14) \ar_{\pi^m \hskip 8pt}@{-} +(4,-8)
\POS(2,6) *+{\bullet}*+{   \hskip 15pt  \c}
\POS(2,6) \ar_{f}@{-} +(0,-8)
\endxy \cong 
\xy
\POS(6,14)\ar^{\hskip 8pt \pi^n}@{-}+(-4,-8) 
\POS(-2,14) \ar_{f \hskip 8pt}@{-} +(4,-8)
\POS(2,6) *+{\bullet}*+{   \hskip 15pt  \c}
\POS(2,6) \ar_{f}@{-} +(0,-8)
\endxy \cong 
\xy
\POS(2,6) \ar_{f}@{-} +(0,8)
\POS(2,6) *+{\bullet}*+{   \hskip 15pt  \c}
\POS(2,6) \ar_{f}@{-} +(0,-8)
\endxy
\end{equation}
As in the previous case, one can associate to this
relation a Hopf ideal in $\cH^{nc}_{{\rm flow},\cP}$.
However, in this case, the relation involves not only
a property of the elementary operations that label the
vertices but also of a specific choice of input functions,
hence it does not define a Hopf ideal in $\cH^{nc}_{{\rm flow},\cV}$.

\smallskip

Thus, from the point of view of computation and algorithms, 
the difference between working with flow charts or with
templates corresponds to two different notions of algorithms.
In the first case algorithms are equivalence classes of
flow charts under relations that can involve both properties of
the elementary operations that hold for arbitrary input functions
and properties that hold for specific input functions, while in
the case of templates, one only allows the first type of relations
in the definition of algorithms.

\smallskip
\subsection{Symmetries: a physics motivated perspective} 
We will be discussing, especially in relation to
Dyson--Schwinger equations, notions of {\em symmetry} in the context
of Hopf algebras. The main motivating example for the notion of symmetry
we adopt is again coming from the setting of quantum field theory, where
gauge symmetries are implemented at the quantum level in the form
of Ward (or Slavonov--Taylor) identities. It was shown by van Suijlekom in
\cite{WvS} that the Slavonov--Taylor identities define a Hopf ideal in the
Connes--Kreimer Hopf algebra of Feynman graphs. It is therefore a natural
approach from the physics perspective to require that a good notion of
symmetry in a setting where the main objects of interest (Feynman
graphs, flow charts, etc.) are described in terms of a Hopf algebra should
be expressed in terms of Hopf ideals. We adopt this viewpoint in the
following, in \S \ref{YanoSec}, where we propose a different approach to
Yanofsky's Galois theory of algorithms, using Hopf ideals in the Hopf algebra
of flow charts as a good notion of equivalence relations defining intermediate
levels (algorithms) between flow charts and primitive recursive functions,
and in \S \ref{DSidealsSec}, where we consider Hopf ideals associated to
solutions of Dyson--Schwinger equations.

\medskip

\section{Dyson--Schwinger equations in the Hopf algebra of flow charts}\label{DSflowSec}

Recall here the formulation of combinatorial Dyson--Schwinger equations in
Hopf algebras of decorated rooted trees, following \cite{BergKr}, \cite{Yeats} and 
especially \cite{Foissy1}, \cite{Foissy2}, \cite{Foissy3}. We then focus on the specific
case of the Hopf algebras of flow charts described above.

\subsection{Insertion operators for decorated planar rooted trees}

We work here with the Hopf algebra of flow charts $\cH^{nc}_{{\rm flow},\cV'}$.
By analogy with the setting of perturbative renormalization \cite{BergKr}, \cite{Yeats},
we define, for each type of elementary operation $\delta\in \{ \b, \c, \r \}$, 
a grafting operator $B^+_\delta$ defined in the following way. Given a monomial $T$ in
$\cH^{nc}_{{\rm flow},\cV'}$, which consists of a forest with vertex labeling by elementary
operations, we define $B^+_\delta(T)$ as the sum of planar graphs obtained by
adding a new root vertex with a number of incoming flags equal to the number
of trees in $T$ and a single output flag, with the vertex decorated by $\delta$.

\subsubsection{Hochschild 1-cocyles}

In the case of perturbative renormalization, it is well known \cite{BergKr} that
the analogous grafting operators $B^+_\delta$, with vertex decorations $\delta$, 
define Hochschild 1-cocycles. 

The 1-cocycle condition for the operator $B^+_\delta$ then consists of the
property that
\begin{equation}\label{cocycle}
 \Delta B^+_\delta = (id \otimes B^+_\delta) \Delta + B^+_\delta \otimes 1. 
\end{equation} 

\begin{lem}\label{1cocycle1}
The operators $B^+_\delta$ on $\cH^{nc}_{{\rm flow},\cV'}$, with $\delta\in \{ \b, \c, \r \}$, satisfy 
the 1-cocycle condition \eqref{cocycle}. 
\end{lem}

\proof This case works exactly as in the renormalization setting, see \cite{BergKr}:
we reproduce the argument here for the reader's convenience.
For an element $x$ in a Hopf algebra $\cH$ with coproduct
$$ \Delta(x) = 1\otimes x + x\otimes 1+ \sum x'\otimes x'', $$
where $x'$ and $x''$ denote lower degree terms,  we use the notation 
$$ \tilde\Delta(x) :=\sum x'\otimes x''. $$
One then sees easily that the cocycle condition \eqref{cocycle} is equivalent to
\begin{equation}\label{Bplustilde}
 \tilde\Delta B^+_\delta = (id \otimes B^+_\delta) \tilde\Delta + id \otimes B^+_\delta(1), 
\end{equation} 
where, as in the renormalization case, with $B^+_\delta(1)=v_\delta$ is a single vertex 
with assigned label $\delta$. The argument is then as in
the original case, with the first term in the right hand side of \eqref{Bplustilde} accounting
for the admissible cuts where the root vertex remains attached to the $\rho_C(T)$
part of an admissible cut of $T$, and the second term counts the case where the admissible
cut separates the root vertex completely.
\endproof

\smallskip

Notice that, for the cocycle condition to hold, we need to be able to assign the same
type of label ($\b$, $\c$, or $\r$) to vertices of arbitrary valence, hence the reason for
including in the set of vertex decorations all the $k$-ary versions of composition 
and recursion and working with the version $\cH^{nc}_{{\rm flow},\cV'}$ of the Hopf algebra
of flow charts.

\smallskip

It is possible to define other interesting grafting operators in the
Hopf algebras of flow charts, which, however, do not satisfy the
cocycle condition. These can still be used to construct Dyson--Schwinger
equations, but the set of solutions does not define a Hopf subalgebra.
This will be discussed in an example in the following subsection.

\smallskip

\subsection{Systems of combinatorial Dyson--Schwinger equations}

Given a graded Hopf algebra $\cH$, we consider the direct product 
$\overline{\cH} =\prod_{n=0}^\infty \cH_n$, whose elements we write
as infinite sums $x=\sum_{n=0}^\infty x_n$, with $x_n\in \cH_n$, endowed
with the associative product and coproduct induced by those of $\cH$.
Following \cite{Foissy1}, (see also \cite{BergKr}, \cite{Yeats})
given a formal power series 
$$ P(t)= \sum_{k=0}^\infty a_k t^k , \ \  \text{ with } \  a_0=1, $$
and a Hochschild 1-cocycle $B^+$ on the Hopf algebra $\cH$,
the associated Dyson--Schwinger equation is given by
\begin{equation}\label{DSP}
X = B^+(P(X)).
\end{equation}
One considers this as an equation in $\overline{\cH}$ and
interprets the infinite sum $P(X)=\sum_k a_k X^k$ as an 
element in $\overline{\cH}$.  

\smallskip

One should read \eqref{DSP} as a fixed point equation for the
nonlinear transformation
$$ X \mapsto B^+(P(X)) $$
of $\overline{\cH}$. Solutions to the fixed point equation are
elements of $\overline{\cH}$ that exhibit a self-similarity property
with respect to the operation of first applying $X\mapsto P(X)$
(in the case of a polynomial, this operation heuristically replaces 
$X$ by a sum of multiple copies of itself weighted by the coefficients 
of the polynomial) and then grafting them together according to 
the cocycle $B^+$. 

\smallskip

In fact, the equation \eqref{DSP} has a unique solution given by
an element $X=\sum_k x_k$ with $x_k\in \cH_k$ whose homogeneous
components are determined inductively by the procedure
\begin{equation}\label{DSsol}
x_{n+1} = \sum_{k=1}^n \sum_{j_1+\cdots+ j_k=n} a_k \, B^+(x_{j_1} \cdots x_{j_k}),
\end{equation}
starting with $x_1$ given by $B^+(1)$, see Lemma 2 of \cite{BergKr} and
Proposition 2 of \cite{Foissy1}.

\smallskip

Notice that we can still define the Dyson--Schwinger equation \eqref{DSP} for
other types of grafting operators that do not necessarily satisfy the 1-cocycle
condition \eqref{cocycle}, such as the examples discussed in the previous
subsection. The existence and uniqueness of solutions, of the form \eqref{DSsol}
is also still valid. Where the 1-cocycle condition is crucially used is in showing
that the associative subalgebra of $\cH$ generated by the components $x_k$
of the solution is also a Hopf subalgebra, see Theorem 3 of \cite{BergKr}.

\subsubsection{Systems of Dyson--Schwinger equations}

As discussed in the work of Foissy \cite{Foissy2}, in cases like our $\cH^{nc}_{{\rm flow},\cV'}$,
where one has different vertex labels $\delta \in \{ \b, \c, \r \}$, one can consider more
complicated {\em systems} of Dyson--Schwinger equations, involving
the grafting operators $B^+_\delta$ on $\cH^{nc}_{{\rm flow},\cV'}$.

\smallskip

For $\delta\in \{ \b, \c, r \}$, we consider the grafting cocycles $B^+_\delta$ on the
Hopf algebra $\cH^{nc}_{{\rm flow},\cV'}$, as above. We also consider data of
three non-constant formal power series in three variables $X=(X_\delta)$
$$ F_\delta(X) = \sum_{k_1,k_2,k_3} a^{(\delta)}_{k_1,k_2,k_3} X_\b^{k_1} X_\c^{k_2} X_\r^{k_3}. $$
To these data we associate the system of Dyson--Schwinger equations as in \cite{Foissy2}
\begin{equation}\label{DSsystem}
X_\delta = B^+_\delta(F_\delta(X)).
\end{equation}

As shown in Proposition 5 of \cite{Foissy2}, these systems of equations also have
a unique solution of the form $X_\delta = \sum_\tau x_\tau \, \tau$, with the sum over all
planar rooted trees with root decorated by $\delta$, with coefficients
\begin{equation}\label{xtaueq}
 x_\tau= (\prod_{k=1}^3 \frac{(\sum_{l=1}^{m_k} p_{\delta,l})!}{\prod_{l=1}^{m_k} p_{\delta,l}!})
a^{(\delta)}_{\sum_{k=1}^3 p_{1,k}, \sum_{k=1}^3 p_{2,k}, \sum_{k=1}^3 p_{3,k}} x^{p_{1,1}}_{\tau_{1,1}}\cdots x_{\tau_{3,m_3}}^{p_{3,m_3}}, 
\end{equation}
when $$\tau =B^+(\tau_{1,1}^{p_{1,1}}\cdots  \tau_{1,m_1}^{p_{1,m_1}} \cdots  \tau_{3,1}^{p_{3,1}}  \cdots \tau_{3,m_3}^{p_{3,m_3}}).$$

\subsection{Dyson--Schwinger solutions: a computational perspective}

The solution $X_\delta = \sum_\tau x_\tau \tau$ of a Dyson--Schwinger equation
$X_\delta = B_\delta^+(F_\delta(X))$ as above, is a formal infinite linear combination
of templates $\tau$, with coefficients $x_\tau$ given by recursively defined computable
functions of the coefficients $a^{(\delta)}_{k_1,k_2,k_3}$ of $F_\delta$. It takes an 
input $f^{\rm input}=(f^{\rm input}_\tau)_\tau$ consisting of a sequence of primitive 
recursive functions, where $f^{\rm input}_\tau=(f^{\rm input}_{\tau,1}, \ldots, 
f^{\rm input}_{\tau,n_\tau})$ are the $n_\tau$ inputs of a template $\tau$ given 
by a tree with $n_\tau$ leaves. It produces an output given by a formal linear
combination $\sum_\tau x_\tau f^{\rm output}_\tau$, where $f^{\rm output}_\tau$
is the output of the template $\tau$ for the given input $f^{\rm input}_\tau$. More
precisely, to avoid dealing with infinite linear combinations, one can consider
the individual components of the solution obtained by recursively solving the
Dyson--Schwinger equation.

\smallskip

For simplicity, we look at how this works in a toy-model example 
of Dyson--Schwinger equation. Let us restrict to only one possible type
of vertex labeling by composition $\c$, and consider an equation of the form
$X = 1+ B^+_{\c} (X^2)$. The components $x_n$ of the solution are
then obtained recursively as
$$ x_0 =1, \ \ \  x_1 = \xy 
\POS(0,4) \ar_{}@{-} +(0,-4)
\POS(0,0) *+{\bullet}*+{   \hskip 15pt  \c} 
\POS(0,0) \ar_{}@{-} +(0,-4)
\endxy, \ \ \
x_2 = \xy 
\POS(0,6) \ar_{}@{-} +(0,-4)
\POS(0,2) *+{\bullet}*+{   \hskip 15pt  \c}
\POS(0,2) \ar_{}@{-} +(0,-4)
\POS(0,-2) *+{\bullet}*+{   \hskip 15pt  \c}
\POS(0,-2) \ar_{}@{-} +(0,-4)
\endxy, \ \ \
$$
$$ x_3 = \xy 
\POS(-2,4) \ar_{}@{-} +(0,-4)
\POS(2,4) \ar_{}@{-} +(0,-4)
\POS(2,0) *+{\bullet}*+{   \hskip 15pt  \c}
\POS(-2,0) *+{\bullet}*-{     \c \hskip 15pt}
\POS(2,0) \ar_{}@{-} +(-2,-4)
\POS(-2,0) \ar_{}@{-} +(2,-4)
\POS(0,-4) *+{\bullet}*+{   \hskip 15pt  \c}
\POS(0,-4) \ar_{}@{-} +(0,-4)
\endxy \ 
+ 4 \xy 
\POS(0,8) \ar_{}@{-} +(0,-4)
\POS(0,4) *+{\bullet}*+{   \hskip 15pt  \c}
\POS(0,4) \ar_{}@{-} +(0,-4)
\POS(0,0) *+{\bullet}*+{   \hskip 15pt  \c}
\POS(0,0) \ar_{}@{-} +(0,-4)
\POS(0,-4) *+{\bullet}*+{   \hskip 15pt  \c}
\POS(0,-4) \ar_{}@{-} +(0,-4)
\endxy, \ \ \
x_4 = 4 \ \ \xy
\POS(-2,8) \ar_{}@{-} +(0,-4)
\POS(-2,4) *+{\bullet}*- {   \c   \hskip 15pt}
\POS(-2,4) \ar_{}@{-} +(0,-4)
\POS(2,4) \ar_{}@{-} +(0,-4)
\POS(2,0) *+{\bullet}*+{   \hskip 15pt  \c}
\POS(-2,0) *+{\bullet}*-{     \c \hskip 15pt}
\POS(2,0) \ar_{}@{-} +(-2,-4)
\POS(-2,0) \ar_{}@{-} +(2,-4)
\POS(0,-4) *+{\bullet}*+{   \hskip 15pt  \c}
\POS(0,-4) \ar_{}@{-} +(0,-4)
\endxy \
+ 2 \ \ \xy 
\POS(-2,4) \ar_{}@{-} +(0,-4)
\POS(2,4) \ar_{}@{-} +(0,-4)
\POS(2,0) *+{\bullet}*+{   \hskip 15pt  \c}
\POS(-2,0) *+{\bullet}*-{     \c \hskip 15pt}
\POS(2,0) \ar_{}@{-} +(-2,-4)
\POS(-2,0) \ar_{}@{-} +(2,-4)
\POS(0,-4) *+{\bullet}*+{   \hskip 15pt  \c}
\POS(0,-4) \ar_{}@{-} +(0,-4)
\POS(0,-8) *+{\bullet}*+{   \hskip 15pt  \c}
\POS(0,-8) \ar_{}@{-} +(0,-4)
\endxy \
+ 8 \xy 
\POS(0,10) \ar_{}@{-} +(0,-4)
\POS(0,6) *+{\bullet}*+{   \hskip 15pt  \c}
\POS(0,6) \ar_{}@{-} +(0,-4)
\POS(0,2) *+{\bullet}*+{   \hskip 15pt  \c}
\POS(0,2) \ar_{}@{-} +(0,-4)
\POS(0,-2) *+{\bullet}*+{   \hskip 15pt  \c}
\POS(0,-2) \ar_{}@{-} +(0,-4)
\POS(0,-6) *+{\bullet}*+{   \hskip 15pt  \c}
\POS(0,-6) \ar_{}@{-} +(0,-4)
\POS(0,10) \ar_{}@{-} +(0,-4)
\endxy, \ \ \
$$
and so on. More realistically, one would have to count also templates with
other possible labelings of vertices besides $\c$, but for simplicity let us
focus on this case. Then, for example, we interpret $x_3$ as a template that,
for a given input $(f_1, f_2)$, consisting of a pair of primitive recursive
functions computes the output function $6 f_1 \circ f_2 + 8 f_1$. Notice that
the self-similar structure of the Dyson--Schwinger equation is not apparent
when one looks at one individual component of the solution but only in the
overall recursive relation $x_{n+1}=\sum_{k=0}^n B_\c^+(x_k x_{n-k})$.

\subsection{Hopf subalgebras and Hopf ideals}

Bergbauer and Kreimer showed that, under the assumption that $B^+_\delta$
is a Hochschild cocycle, solutions of Dyson--Schwinger equations determine a sub-Hopf
algebra. Namely, they considered Dyson--Schwinger
equation in a Hopf algebra $\cH$ of the form
\begin{equation}\label{BergKrDS}
 X = 1 + \sum_{n=1}^\infty c_n \, B^+_\delta (X^{n+1}). 
\end{equation} 
One then considers the associative algebra $\cA$ defined as the 
{\em subalgebra} of $\cH$ generated by the components $x_n$,
with $n\geq 0$ of the unique solution of this Dyson--Schwinger equations.
In Theorem 3 of \cite{BergKr}, they showed that $\cA$ is in fact a {\em sub-Hopf}
algebra, by inductively using the cocycle condition \eqref{cocycle} to see that
\begin{equation}\label{BKdelta}
 \Delta(x_n) = \sum_{k=0}^n \Pi^n_k \otimes x_k, \ \ \text{ 
where } \ \  \Pi^n_k = \sum_{j_1+\cdots + j_{k+1}= n-k} x_{j_1} \cdots x_{j_{k+1}}. 
\end{equation}

\smallskip

This result was extended by Foissy to the more general form \eqref{DSP}
of Dyson--Schwinger equations in \cite{Foissy1} and to systems
of Dyson--Schwinger equations in \cite{Foissy2}. In Theorem 4 of \cite{Foissy1}
it is shown that, for an equation of the form \eqref{DSP}, the subalgebra $\cA$
spanned by the solutions is a Hopf subalgebra if and only if the formal series $P(t)$
satisfies the differential equation 
\begin{equation}\label{eqdiff}
(1-\alpha\beta t) P'(t)=\alpha P(t),  \ \ \  \text{ with } \ \ \  P(0)=1,
\end{equation}
for some $\alpha,\beta\in \K$, again under the assumption that $B^+$ satisfies
the cocycle condition \eqref{cocycle}. Similarly, in \cite{Foissy2} combinatorial
conditions and conditions on the multivariable series $F_i$ are
identified that completely characterize when the solutions of that systems of 
Dyson--Schwinger equations generate a Hopf algebra, under the assumption
that the $B^+$ satisfies the cocycle condition \eqref{cocycle}.

\subsubsection{Ideals of Dyson--Schwinger solutions}\label{DSidealsSec}

It is natural to consider not only the associative {\em subalgebra} $\cA$
spanned by the components of the solutions of Dyson--Schwinger
equations, but also the {\em ideal} $\cI$ spanned by the $( x_n)_{n\geq 1}$. 
As in the case of the subalgebra $\cA$, it is also natural to ask for conditions
that will ensure that it is a Hopf subalgebra, so with the ideal $\cI$ of solutions
it is natural to ask for conditions ensuring that it is a Hopf ideal, so that it makes
sense to define a quotient Hopf algebra, playing the role of the ring of functions
on the variety of solutions.

\smallskip

As an example, 
we consider here the Dyson--Schwinger equation \eqref{DSP}, with $B^+=B^+_\b$
the grafting operator to a vertex labelled with the bracketing operation. %
Assuming that parallel computations that do not feed into each other can be
inputed in arbitrary order at the vertex, we can regard this grafting operator as descending 
to the commutative quotient $\cH_{{\rm flow}, \cV'}$ 
of the noncommutative Hopf algebra $\cH^{nc}_{{\rm flow}, \cV'}$. 
This heuristic rationale for passing to the commutative quotient does not take into account
a proper definition of a notion of parallel processes, which would require considering the
notion of leveled trees and, in the binary case, Hopf algebra structures related to 
permutohedra instead of associahedra. We refer the reader to 
\cite{FoLaSo}, \cite{Holt2}, \cite{LoRo}. %

\smallskip

Let $X=\sum_{n\geq 1} x_n$ be the unique solution of \eqref{DSP}, obtained as in \eqref{DSsol},
with $x_1=B^+_\b(1)= v_\b$ a single vertex labeled with the $\b$ operation. 
We define $\cI$ to be the ideal in the commutative algebra $\cH_{{\rm flow}, \cV'}$
generated by the $x_n$ with $n\geq 1$.

\smallskip

The same conditions that ensures that $\cA$ is a Hopf subalgebra also ensure
that $\cI$ is a Hopf ideal. We verify it in the case of \eqref{BergKrDS}. The 
more general case of \cite{Foissy1} is similar.

\smallskip

\begin{lem}\label{BKideal}
The ideal $\cI$ generated by the components $x_n$ with $n\geq 1$ of the
solution of \eqref{BergKrDS} is a Hopf ideal.
\end{lem}

\proof Elements of the ideal $\cI$ in $\cH=\cH_{{\rm flow}, \cV'}$ are finite
sums $\sum_{m=1}^M h_m x_{k_m}$, with $h_m \in \cH$ and $x_k$
the coordinates of the unique solution of \eqref{BergKrDS}.  The condition
that $\cI$ is a Hopf ideal is that $\Delta(\cI)\subset  \cI \otimes \cH \oplus \cH \otimes \cI$.
Using the formula \eqref{BKdelta} for the coproduct of the elements $x_k$, we
see that we obtain a sum of terms of which the primitive part
$1\otimes x_k + x_k \otimes 1$ is in $\cH \otimes \cI \oplus \cI \otimes \cH$ and
all the other terms $\tilde\Delta(x_k)$ are in $\cI \otimes \cI$. Then the
coproducts $\Delta(h_m x_{k_m})$ will also be in $\cH \otimes \cI \oplus \cI \otimes \cH$.
\endproof

Notice that we can also work with the noncommutative Hopf algebra 
$\cH^{nc}_{{\rm flow}, \cV'}$ and the two sided ideal with elements of
the form $\sum_m h_m x_{k_m} \ell_m$, with $h_m, \ell_m \in \cH^{nc}_{{\rm flow}, \cV'}$,
but for the considerations that follow (see \S \ref{YanoSec} below) it is more
natural to work with the commutative quotient.

\medskip

\subsection{Yanofsky's Galois theory of algorithms}\label{YanoSec}

As we have seen above, one can pass to commutative quotients $\cH_{\rm flow}$ of
the noncommutative Hopf algebras of flow charts. 
The meaning of passing to the commutative quotient $\cH_{\rm flow}$ 
is best expressed as follows: one can think of monomials
in $\cH_{\rm flow}^{nc}$  as diagrams for parallel computations
of a certain number of outputs (one for each planar tree in the forest). Imposing commutativity
then corresponds to the reasonable expectation that if one has two parallel computations 
that do not feed one into the other, then computing them in either order will not affect the result.
However, when one introduces operations that graft the
output of one three to the input of another, with $\c$ and $\r$ labels at the vertices, 
imposing the commutativity relation is no longer appropriate, as the result of 
grafting would not be well defined as a {\em planar} tree. In the case of grafting with
$\c$ label, we can still pass to the commutative quotient, as argued above.

\smallskip

\subsubsection{Commutative Hopf algebras and affine group schemes}

An advantage of working with commutative Hopf algebras is that they are
dual to affine group schemes. Thus, the Hopf algebra $\cH_{\rm flow}$
determines an affine group scheme $G_{{\rm flow}}$ such that, for any 
commutative $\K$-algebra $A$, one obtains a group $G_{{\rm flow}}(A)={\rm Hom}(\cH_{\rm flow},A)$,
the set of homomorphisms of $\K$-algebras, with the product $(\phi_1\star \phi_2)(x)
=\langle \phi_1\otimes \phi_2, \Delta(x)\rangle$ dual to the coproduct and the inverse
determined by the antipode. 

\medskip

\subsubsection{Galois groups of algorithms}

The idea elaborated by Yanofsky in the recent work \cite{Yano2}, of a Galois theory
of algorithms, aims at considering all possible ``reasonable sets of relations" on
programs that would correspond to ``implementing
the same algorithm", viewing ``algorithms" as an intermediate level between the
labelled planar rooted trees and the primitive recursive functions they compute.
Our point of view here is similar and directly inspired by the approach proposed by
Yanofsky, with the main difference that we seek to implement relations at the level
of the Hopf algebra $\cH_{\rm flow}^{nc}$ of flow charts.

\smallskip

The automorphism groups that arise in Yanofsky's setting are typically products
of symmetric groups and the notion of equivalence relations considered does
not take into account any additional structure on flow charts, such as Manin's
Hopf algebra point of view. Moreover, a possible drawback of the current
formulation of the approach of \cite{Yano2} lies in the fact that different symmetry
groups can have the same orbit space and give rise to the same equivalence
on algorithms. We propose instead to consider a different approach
where equivalence relations on algorithms are imposed at the level of the 
Hopf algebra of flow charts.

\smallskip

Our proposal for an alternative approach to a Galois theory of algorithms is to take
into account the Hopf algebra structure as part of the construction. By this we mean
defining as ``reasonable relations" between the planar rooted trees of 
$\cH_{\rm flow}^{nc}$ those that correspond to passing to quotient Hopf algebras. 
Each such quotient is determined by 
a Hopf ideal $\cI$ in $\cH_{\rm flow}^{nc}$, the kernel of the map to the quotient Hopf 
algebra $\cH_\cI$. 

\smallskip

In the commutative setting $\cH_{\rm flow}$ one can view, dually, the quotient Hopf algebras $\cH_I$
as sub-group schemes $G_I$ of the affine group scheme of characters $G_{{\rm flow}}$. The group
schemes $G_I$ obtained in this way replace, in our approach, the basic objects of 
interest in Yanofsky's Galois theory of algorithms.

\medskip

\subsubsection{Hopf ideals and Galois groups}

In light of Yanofsky's point of view, %
it is natural to consider ideals in $\cH_{\rm flow}$ generated by 
solutions of Dyson--Schwinger equations, and the resulting quotient 
Hopf algebras, rather than sub-Hopf algebras. The Hopf ideals
and quotient Hopf algebras define, dually, the affine group schemes
that provide our propose notion of Galois groups of algorithms.

\smallskip

As a heuristic geometric interpretation, in the Hopf algebraic setting, 
we can think of the quotient $\cH/\cI$ of a Hopf algebra by a Hopf ideal 
as an analog of the algebro-geometric quotient $R/I$ of a polynomial 
ring $R$ by a prime ideal $I$, which provides the coordinate ring of the 
affine variety $V$ defined by $I$. In the case of the Hopf ideal $\cI$
determined by the components of the solution of the Dyson--Schwinger
equation, we can therefore think of $\cH/\cI$ as 
the ``ring of function" of the variety of solutions. The associated group
scheme $G_I$ consists of those symmetries in $G_{{\rm flow}}$ that are
compatible with the self-similarity structure described by the 
Dyson--Schwinger equation.

\smallskip

\begin{prop}\label{SDgroup}
A Dyson--Schwinger equation \eqref{BergKrDS} in the 
commutative Hopf algebra $\cH_{\rm flow}$ determines
a Galois group $G_{P} \subset G_{{\rm flow}}$.
\end{prop}

This is simply a reformulation of Lemma \ref{BKideal}, with our notion
of what we consider to be an appropriate notion of Galois group. A similar result holds for
Dyson--Schwinger equations \eqref{DSP} with \eqref{eqdiff}. While, on the one hand,
our point of view takes into account more information, coming from the Hopf algebra
structure, at another level it is less structured than Yanofsky's setting: in fact, we work
here with the Hopf algebra where flow charts only have vertex decorations, so that
the input functions (and the corresponding edge labelings through the graph) are
not specified, while in Yanofsky's setting the underlying functions are assigned and
preserved by the symmetries.

\medskip

\subsection{A non-Hopf example with binary trees}

In the setting of Dyson--Schwinger equations described above, it is
crucial that we consider the planar rooted trees in $\cH^{nc}_{{\rm flow}, \cV'}$
without external edges. In fact, this allows us to define the grafting operators
$B^+_\delta$ as grafting any number of trees to the {\em same} root consisting
of a single vertex decorated by $\delta$. If we reintroduce external edges, then
the graphs consisting of a single vertex but with different numbers of incoming
edges are viewed as different generators in the Hopf algebra and they no
longer define a cocycle $B^+_\delta$.

\smallskip

To see this more explicitly, we focus on the case of binary trees, keeping track of
external edges.
Consider the following primitive elements in the Hopf algebra $\cH^{nc,0}_{{\rm flow}, \cV}$:
\begin{equation}\label{prim3}
\xy
\POS(6,14) \ar_{}@{-}+(-4,-8)
\POS(-2,14) \ar_{}@{-} +(4,-8)
\POS(2,6) *+{\bullet}*+{   \hskip 15pt  \r}
\POS(2,6) \ar_{}@{-} +(0,-8)
\endxy \hskip 10pt , \hskip 10pt 
\xy
\POS(6,14) \ar_{}@{-}+(-4,-8)
\POS(-2,14) \ar_{}@{-} +(4,-8)
\POS(2,6) *+{\bullet}*+{   \hskip 15pt  \b} 
\POS(2,6) \ar_{}@{-} +(0,-8)
\endxy 
\hskip 10pt , \hskip 10pt
\xy
\POS(6,14) \ar_{}@{-}+(-4,-8)
\POS(-2,14) \ar_{}@{-} +(4,-8)
\POS(2,6) *+{\bullet}*+{   \hskip 15pt  \c}
\POS(2,6) \ar_{}@{-} +(0,-8)
\endxy
\end{equation}

For each of these graphs $\tau$ one can define a grafting operation $B^+_\tau$
by setting the result equal to zero on monomials $T$ that cannot be grafted to $\tau$.
This $B^+_\tau$ clearly does {\em not} satisfy the cocycle condition because 
one would have to have $B^+_\tau(1)=0$, but then the expressions 
$\tilde\Delta B^+_\tau$ and $(id \otimes B^+_\tau) \tilde\Delta$, when applied
to elements $T=\tau_1\amalg \tau_2$ that are not in the kernel, would differ by
a term corresponding to the cut that separates $\tau$ from $T$. On the other
hand, if $B^+_\tau(1)=\tau$ as this case would require, then the cocycle condition 
fails on elements in the kernel of $B^+_\tau$. 

\smallskip

Another way of defining a grafting operator $B^+_\tau$ with $\tau$ one of 
the trees \eqref{prim3} is to define, for a monomial $T=\tau_1\amalg\cdots \amalg \tau_n$ the
grafting $B^+_\tau(T)$ to be obtained by just gluing one of the outputs of $T$ (say, the first one)
to one of the two inputs of $\tau$ (say, the first one), %
with the result of grafting taken to be $\tau$ itself if $T=\emptyset$. %
We use this choice in the explicit
example below. Again, however, one sees that the grafting defined in this way does {\em not}
satisfy the cocycle condition. In fact, consider the case of a monomial $T$ consisting
of a forest with more than one component. Then only one of the components of $T$ is
glued to $\tau$ by $B^+_\tau$. This means that, in the computation of the term
$\tilde\Delta B^+_\tau$ there is not just one but several admissible cuts that separate
$\tau$ from $T$, namely all the admissible cuts that include a cut of the edge used
for grafting together with an arbitrary cut of the remaining components not connected to $\tau$.
Thus, the difference between $\tilde\Delta B^+_\tau$ and $(id \otimes B^+_\tau) \tilde\Delta$
does not consist only of the term $id \otimes B^+_\delta(1)=id \otimes \tau$.

\smallskip

However, one can still consider a Dyson--Schwinger equation of the
form
$$ X = 1 + \sum_\delta B^+_{\tau_\delta}(X^2), $$
where the sum is on the three labels $\delta \in \{ \b, \c, \r \}$ of the vertex of the tree $\tau_\delta$
as in \eqref{prim3}.
This admits an explicit solution, as in the general case above, which
are of the form
$$ x_{n+1} = \sum_{k=0}^n \sum_\delta B^+_{\tau_\delta}(x_k x_{n-k}). $$
In $\overline{\cH}^{0}_{{\rm flow}, \cV}$, one can see that the first few 
terms are of the form $x_0 = I$ and

$$x_1 = 
\xy
\POS(6,14) \ar_{}@{-}+(-4,-8)
\POS(-2,14) \ar_{}@{-} +(4,-8)
\POS(2,6) *+{\bullet}*+{   \hskip 15pt  \textbf{r}}
\POS(2,6) \ar_{}@{-} +(0,-8)
\endxy \hskip 10pt + \hskip 10pt 
\xy
\POS(6,14) \ar_{}@{-}+(-4,-8)
\POS(-2,14) \ar_{}@{-} +(4,-8)
\POS(2,6) *+{\bullet}*+{   \hskip 15pt  \textbf{b}} 
\POS(2,6) \ar_{}@{-} +(0,-8)
\endxy
\hskip 10pt + \hskip 10pt 
\xy
\POS(6,14) \ar_{}@{-}+(-4,-8)
\POS(-2,14) \ar_{}@{-} +(4,-8)
\POS(2,6) *+{\bullet}*+{   \hskip 15pt  \textbf{c}}
\POS(2,6) \ar_{}@{-} +(0,-8)
\endxy$$ 

$$x_2 = \sum_a B_+^a ( 2x_1 x_0 )$$
 $$ = 2  
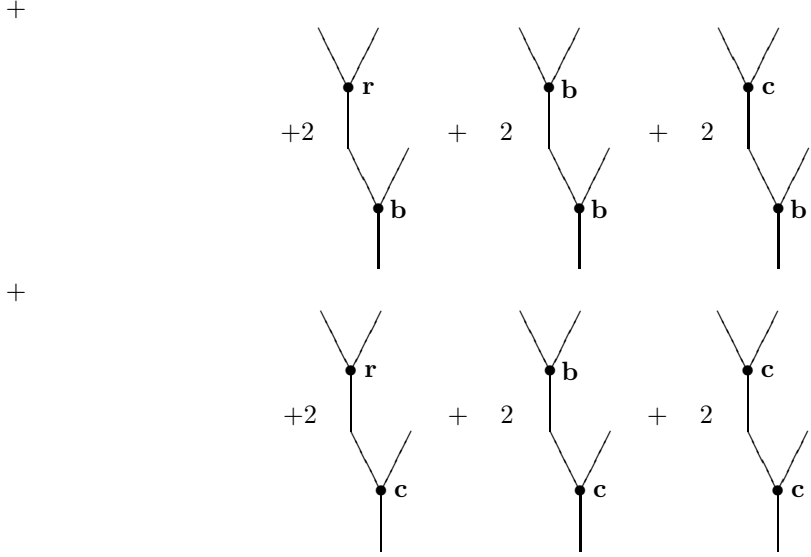
\xy
\POS(6,14) \ar_{}@{-}+(-4,-8)
\POS(-2,14) \ar_{}@{-} +(4,-8)
\POS(2,6) *+{\bullet}*+{   \hskip 15pt  \textbf{r}}
\POS(2,6) \ar_{}@{-} +(0,-8)
\POS(2,-2) \ar_{}@{-} +(4,-8)
\POS(10,-2) \ar_{}@{-}+(-4,-8)
\POS(6,-10) *+{\bullet}*+{   \hskip 15pt  \textbf{r}}
\POS(6,-10) \ar_{}@{-}+(0,-8)
\endxy
\hskip 10pt + \hskip 10pt
2 \xy
\POS(6,14) \ar_{}@{-}+(-4,-8)
\POS(-2,14) \ar_{}@{-} +(4,-8)
\POS(2,6) *+{\bullet}*+{   \hskip 15pt  \textbf{b}}
\POS(2,6) \ar_{}@{-} +(0,-8)
\POS(2,-2) \ar_{}@{-} +(4,-8)
\POS(10,-2) \ar_{}@{-}+(-4,-8)
\POS(6,-10) *+{\bullet}*+{   \hskip 15pt  \textbf{r}}
\POS(6,-10) \ar_{}@{-}+(0,-8)  \endxy
\hskip 10pt + \hskip 10pt
2 \xy
\POS(6,14) \ar_{}@{-}+(-4,-8)
\POS(-2,14) \ar_{}@{-} +(4,-8)
\POS(2,6) *+{\bullet}*+{   \hskip 15pt  \textbf{c}}
\POS(2,6) \ar_{}@{-} +(0,-8)
\POS(2,-2) \ar_{}@{-} +(4,-8)
\POS(10,-2) \ar_{}@{-}+(-4,-8)
\POS(6,-10) *+{\bullet}*+{   \hskip 15pt  \textbf{r}}
\POS(6,-10) \ar_{}@{-}+(0,-8) 
\endxy$$
\hskip 10pt + \hskip 10pt
$$
+2  \xy
\POS(6,14) \ar_{}@{-}+(-4,-8)
\POS(-2,14) \ar_{}@{-} +(4,-8)
\POS(2,6) *+{\bullet}*+{   \hskip 15pt  \textbf{r}}
\POS(2,6) \ar_{}@{-} +(0,-8)
\POS(2,-2) \ar_{}@{-} +(4,-8)
\POS(10,-2) \ar_{}@{-}+(-4,-8)
\POS(6,-10) *+{\bullet}*+{   \hskip 15pt  \textbf{b}}
\POS(6,-10) \ar_{}@{-}+(0,-8)
\endxy
\hskip 10pt + \hskip 10pt
2 \xy
\POS(6,14) \ar_{}@{-}+(-4,-8)
\POS(-2,14) \ar_{}@{-} +(4,-8)
\POS(2,6) *+{\bullet}*+{   \hskip 15pt  \textbf{b}}
\POS(2,6) \ar_{}@{-} +(0,-8)
\POS(2,-2) \ar_{}@{-} +(4,-8)
\POS(10,-2) \ar_{}@{-}+(-4,-8)
\POS(6,-10) *+{\bullet}*+{   \hskip 15pt  \textbf{b}}
\POS(6,-10) \ar_{}@{-}+(0,-8)  \endxy
\hskip 10pt + \hskip 10pt
2 \xy
\POS(6,14) \ar_{}@{-}+(-4,-8)
\POS(-2,14) \ar_{}@{-} +(4,-8)
\POS(2,6) *+{\bullet}*+{   \hskip 15pt  \textbf{c}}
\POS(2,6) \ar_{}@{-} +(0,-8)
\POS(2,-2) \ar_{}@{-} +(4,-8)
\POS(10,-2) \ar_{}@{-}+(-4,-8)
\POS(6,-10) *+{\bullet}*+{   \hskip 15pt  \textbf{b}}
\POS(6,-10) \ar_{}@{-}+(0,-8) 
\endxy$$
\hskip 10pt + \hskip 10pt
$$+2  \xy
\POS(6,14) \ar_{}@{-}+(-4,-8)
\POS(-2,14) \ar_{}@{-} +(4,-8)
\POS(2,6) *+{\bullet}*+{   \hskip 15pt  \textbf{r}}
\POS(2,6) \ar_{}@{-} +(0,-8)
\POS(2,-2) \ar_{}@{-} +(4,-8)
\POS(10,-2) \ar_{}@{-}+(-4,-8)
\POS(6,-10) *+{\bullet}*+{   \hskip 15pt  \textbf{c}}
\POS(6,-10) \ar_{}@{-}+(0,-8)
\endxy
\hskip 10pt + \hskip 10pt
2 \xy
\POS(6,14) \ar_{}@{-}+(-4,-8)
\POS(-2,14) \ar_{}@{-} +(4,-8)
\POS(2,6) *+{\bullet}*+{   \hskip 15pt  \textbf{b}}
\POS(2,6) \ar_{}@{-} +(0,-8)
\POS(2,-2) \ar_{}@{-} +(4,-8)
\POS(10,-2) \ar_{}@{-}+(-4,-8)
\POS(6,-10) *+{\bullet}*+{   \hskip 15pt  \textbf{c}}
\POS(6,-10) \ar_{}@{-}+(0,-8)  \endxy
\hskip 10pt + \hskip 10pt
2 \xy
\POS(6,14) \ar_{}@{-}+(-4,-8)
\POS(-2,14) \ar_{}@{-} +(4,-8)
\POS(2,6) *+{\bullet}*+{   \hskip 15pt  \textbf{c}}
\POS(2,6) \ar_{}@{-} +(0,-8)
\POS(2,-2) \ar_{}@{-} +(4,-8)
\POS(10,-2) \ar_{}@{-}+(-4,-8)
\POS(6,-10) *+{\bullet}*+{   \hskip 15pt  \textbf{c}}
\POS(6,-10) \ar_{}@{-}+(0,-8) 
\endxy$$

One can continue with the successive coefficients, such as 
$x_3 =  \sum_a B_+^a ( x_1^2 + 2x_2 x_0 )$, using the facts that
for a vertex labeled by recursion, the order of inputs always matters, 
as well as for composition, while we can assume that order of
inputs does not matter for bracketing as 
a consequence of the nature of parallel computation.

\smallskip

This example shows that, if we want to work with binary trees with 
external edges, which is a very natural
choice from the point of view of computation theory, and we want to
obtain a grafting operation that is non-trivial on monomials of
arbitrary degree, we would have to extend the grafting 
to include, besides the primitive two-valent trees \eqref{prim3},
also a choice of binary trees with larger numbers of input flags.
This cannot be accommodated in the usual setting of combinatorial
Dyson--Schwinger equations based on Hochschild cocycles in
Hopf algebras. However, we shall see below that such 
generalizations exist naturally when one reformulates
Dyson--Schwinger equations in the operadic setting.

\medskip

\section{Operadic viewpoint}\label{operadSec}

Instead of using a Hopf algebraic setup to describe flow charts and
primitive recursive functions, one can also adopt an operadic viewpoint,
as suggested by Manin in \cite{Man-Zipf}.

\smallskip

Let us consider, as above, planar rooted trees $\tau$ with vertex labeling by
the elementary operations of type $\{ \b, \c, \r \}$, oriented so that each tree has
a single outgoing flag attached to the root vertex and a certain number of
incoming flags. 

\smallskip

\begin{defn}\label{flowoperad}
We define the {\em operad of flow charts} $\cO_{{\rm flow}, \cV'}$ by setting
$\cO(n)$ to the the $\K$-vector space spanned by labelled planar 
rooted trees with $n$ incoming flags and the operad composition
operations
$$ \circ_\cO : \cO(n)\otimes \cO(m_1)\otimes \cdots \otimes \cO(m_n) \to \cO(m_1+\cdots +m_n) $$
are given on generators $\tau \otimes \tau_1\otimes \cdots \otimes \tau_n$ 
by grafting the output flag of the tree $\tau_i$ to the $i$-th input flag of the tree $\tau$.
\end{defn}

\smallskip

As pointed out in \cite{BergKr}, Dyson--Schwinger equations admit a natural operadic
interpretation. Namely, given a formal series $P(t)=1+ \sum_{k=1}^\infty a_k t^k$
and a collection $\beta =( \beta_n )$ with $\beta_n \in \cO(n)$, we consider 
the equation
\begin{equation}\label{DSoperad}
X = \beta (P(X)), 
\end{equation}
where $X= \sum_k x_k$ is a formal sum with $x_k \in \cO(k)$. In  
the right hand side we have $\beta(P(X))_1=1+\beta_1 \circ x_1$, where $1$ is 
the identity in $\cO(1)$, and for $n\geq 2$
\begin{equation}\label{betaPXn}
\beta(P(X))_n = \sum_{k=1}^n   \sum_{j_1+\cdots+ j_k=n} a_k \,\,\, 
\beta_k\circ( x_{j_1} \otimes \cdots \otimes x_{j_k} ), 
\end{equation}
with $x_{j_1} \otimes \cdots \otimes x_{j_k}\in \cO(j_1)\otimes \cdots \otimes \cO(j_k)$, so that
the composition $\beta_k \circ_{\cO} (x_{j_1} \otimes \cdots \otimes x_{j_k}) \in \cO(n)$,
since $j_1+\cdots+ j_k=n$.

\smallskip

\begin{prop}\label{solnDSoperad}
For $\cO=\cO_{{\rm flow}, \cV'}$ the operad of flow charts, 
if $a_1 \beta_1 \neq 1\in \cO(1)$, 
the operadic Dyson--Schwinger equation \eqref{DSoperad} has a unique solution 
$X\in \prod_{n\geq 1} \cO(n)$ given inductively by
\begin{equation}\label{operadSol}
 (1-a_1 \beta_1) \circ x_{n+1} = \sum_{k=2}^{n+1} \sum_{j_1+\cdots j_k =n+1}  a_k \,\,
\beta_k \circ (x_{j_1} \otimes \cdots \otimes x_{j_k}). 
\end{equation}
\end{prop}

\proof First observe that in the operad of flow charts
$\cO(1)$ is one-dimensional, since it is spanned by
the trees with a single vertex, one incoming and one outgoing flag, with
the operations $\b$, $\c$ or $\r$ at the vertex, but in the case of a single
input, all of these operations are the identity, so that elements of $\cO(1)$
are scalar multiples of the identity map, viewed as operations on primitive
recursive functions. Thus, the element $\beta_1 \in \cO(1)$ is a scalar
multiple of the identity and, if we assume that $a_1 \beta_1 \neq 1$ in $\cO(1)$,
then $1- a_1 \beta_1$ is invertible. Since we have  $\beta(P(X))_1=1+\beta_1 \circ x_1$
as the term in $\cO(1)$ in the right hand side of \eqref{DSoperad},  we obtain
$(1- a_1 \beta_1) x_1= 1$, which fixes $x_1 = (1-a_1 \beta_1)^{-1} \in \cO(1)$.
At the next step we have, from \eqref{betaPXn},
$x_2= a_1 \beta_1 \circ x_2  + a_2 \beta_2 \circ (x_1 \otimes x_1)$, which
gives $x_2 = (1-a_1 \beta_1)^{-1} ( a_2 \beta_2 \circ (x_1 \otimes x_1))$, with
$x_1$ as above. At the $(n+1)$-st step, $x_{n+1}$ is then determined uniquely
by \eqref{operadSol} in terms of the coefficients $a_k$ and the elements
$\beta_k$ for $k=1,\ldots, n+1$. 
\endproof

Let then $\cO_{\beta,P}(n)$ denote the $\K$-linear span of all
the compositions $x_k \circ (x_{j_1}\otimes \cdots \otimes x_{j_k})$,
for $k=1, \ldots, n$ and $j_1+\cdots +j_k=n$, with $x_k$ the coordinates
of the solution $X=\sum_k x_k$ of the Dyson--Schwinger equation
determined by $\beta=(\beta_n)$ and $P(t)$. The $\cO_{\beta,P}(n)$
form a sub-operad under the composition maps induced from $\cO(n)$.

\smallskip

For instance, for a choice of one of the operations $\delta\in \{ \b, \c, \r \}$,
we can take $a_1\neq 1$ and the element $\beta_k$ given by
the tree with a single vertex marked by $\delta$, with $k$ incoming
flags and one outgoing flag. This choice gives an operadic reformulation
of the Hopf-theoretic Dyson--Schwinger equations with the cocycles
$B^+_\delta$. However, we can now consider also more general
operadic Dyson--Schwinger equations for different choices of
$\beta=(\beta_n)$ that do not correspond to Hochschild cocycles
in the Hopf algebra setting. 

\smallskip

Restricting to the sub-operad $\cO_{\beta,P}(n)$ generated by solutions
of the Dyson--Schwinger equation should be regarded as considering those
operations (formal combinations of flow charts) that satisfy a self-similarity
property with respect to the transformation $X \mapsto \beta(P(X))$.

\subsubsection{The case of binary trees}

In particular, we can now revisit the example of the binary trees in this operadic
setting. Let $\cO_{{\rm flow}, \cV}^0(n)$ be the $\K$-linear space spanned by
the planar {\em binary} trees with vertex labeling by the binary operations
$\b=\b_2$, $\r$, $\c$. These form a sub-operad of $\cO_{{\rm flow},\cV'}$ with
the induced composition maps. 

For a fixed $\delta \in \{ \b, \r, \c \}$ let $\beta_{1,\delta}$ be the binary tree with a
single vertex, two incoming and one outgoing flag, as in \eqref{prim3}. We then
consider a choice $\beta_n$ of a binary tree with $n$ input flags, for each $n\geq 1$,
with a fixed choice of vertex labelings in $\{ \b, \r, \c \}$, not necessarily all of the same type. 
Each such choice of $\beta=(\beta_n)$, together with a choice of $P(t)$, 
now determines an operadic Dyson--Schwinger equation of the form \eqref{DSoperad},
in the binary setting, with a corresponding solution as in Proposition \ref{solnDSoperad}.

\subsubsection{Systems of Dyson--Schwinger equations in operads}

By analogy to what happens in Hopf algebras, \cite{Foissy2}, one can
pass from the case of a single Dyson--Schwinger equation to systems
of Dyson--Schwinger equations. In the operadic setting a system of
Dyson--Schwinger equations would be determined by the data of formal
series 
$F_\delta(t_1, t_2, t_3)=\sum_{k_1,k_2,k_3} a^{(\delta)}_{k_1,k_2,k_3} t_1^{k_1} t_2^{k_2} t_3^{k_3}$
and of elements $\beta^{(\delta)}=( \beta^{(\delta)}_n )$, where for $\delta\in \{ \b,\c,\r \}$,
the elements $\beta^{(\delta)}_n \in \cO(n)$ are realized by rooted trees with root vertex marked
by $\delta$. The operadic systems of  Dyson--Schwinger equations are then of the form
$$ X_\delta = \beta^{(\delta)} (F_\delta(X)). $$
The explicit form of the solutions is more cumbersome than in the case of a single
equation in Proposition \ref{solnDSoperad} above, and it follows a pattern similar to
the derivation of the explicit solutions of Hopf theoretic systems of Dyson--Schwinger
equations \eqref{DSsystem} as in \cite{Foissy2}.
These systems detect more elaborate forms of self-similarity in the
operad of flow charts, involving different families of grafting operations
simultaneously.

\subsubsection{Operads and Properads}

As mentioned in \cite{Man-Zipf}, one can extend the Hopf algebra of flow
charts by considering, instead of labelled embedded rooted trees, more
general labelled embedded acyclic graphs, namely graphs endowed
with an acyclic orientation. Correspondingly, the operad of flow charts
would be replaced by a properad, where the compositions extend from
grafting output and input flags of trees to grafting outputs and inputs of
acyclic graphs.   

\smallskip

The notion of {\em properad} was introduced in \cite{Val} as an
intermediate notion between operad and prop. Namely, a
properad parameterizes operations with varying numbers of
inputs and outputs that are labelled by connected acyclic graphs,
while the operad case uses trees (varying number of inputs
and a single output) and props allow for disconnected graphs.

\smallskip

More precisely, we consider the properad $\cP_{{\rm flow},\cV'}$
where $\cP(m,n)$ is the $\K$-vector space spanned by planar 
connected directed (acyclic) graphs with $m$ incoming flags 
and $n$ outgoing flags, with vertices decorated by operation that
include the elementary $\b$, $\c$, $\r$,  with $m$ inputs and one output,
but also include now a chosen set of additional operations with $m$ 
inputs and $n$ outputs. As we have seen in the case of
extending the operations $\c$ and $\r$ from binary to $m$-ary
inputs, in this case also such additional operations with $m$ inputs
and $n$ outputs acting on primitive recursive functions can be
decomposed in terms of the elementary operations: the resulting
operations associated to vertices with $m$ incoming and $n$ outoing
flags can be regarded as ``macros", in the sense discussed in  \S 2 of \cite{Yano2}.
The properad has composition operations
$$ \cP(m,n) \otimes \cP(j_1,k_1) \otimes \cdots \otimes \cP(j_\ell, k_\ell) \to
\cP(j_1+\cdots +j_\ell, n), \ \  \text{ for } \ \ 
k_1+\cdots + k_\ell = m. $$

\smallskip

\subsubsection{Dyson--Schwinger equations in properads}

We can define Dyson--Schwinger equations for properads, extending
the operadic case considered above. Namely we consider again a
choice of a formal power series $P(t) =1 +\sum_k a_k t^k$ and a
collection $\beta =(\beta_{m,n})$ of elements with $\beta_{m,n}\in \cP(m,n)$.
We then consider the equation
\begin{equation}\label{DSproperad}
X = \beta(P(X))
\end{equation}
as in \eqref{DSoperad}, where now the right hand side has $(m,n)$-component
in $\cP(m,n)$ given by
\begin{equation}\label{betaPXprop}
\beta(P(X))_{m,n} = \sum_{k=1}^m a_k \sum_{j_1+\ldots j_k=m \atop i_1+\cdots i_k=\ell} 
\beta_{\ell,n}\circ (x_{j_1,i_1}\otimes \cdots \otimes x_{j_k,i_k}).
\end{equation}

\smallskip

In order to construct solutions of the properad Dyson--Schwinger equations, we
introduce transformations $\Lambda_n = \Lambda_n(a,\beta)$ with 
$$ \Lambda_n(a,\beta) : \oplus_{k=1}^n \cP(n,k) \to \oplus_{k=1}^n \cP(n,k), \ \ \ \text{ with } \ \ \
\Lambda_n(a,\beta)_{ij} = a_j  \beta_{j,i} . $$

\smallskip

\begin{thm}\label{thmsolnDSprop}
If for all $n\geq 1$ the transformation
$I - \Lambda_n(a,\beta)$ is invertible, with $I$ the identity on $\oplus_{k=1}^n \cP(n,k)$, 
then the properad Dyson--Schwinger equation
\eqref{DSproperad} admits a unique solution, given by 
$x_{1,1}=\Lambda_1^{-1}$, and for $m<n$ by
\begin{equation}\label{xmlessn}
x_{m,n} = \sum_{k=1}^m a_k \beta_{k,n}\circ \left( \sum_{\ell=1}^k \sum_{j_1+\cdots+ j_\ell =m \atop
i_1+\cdots+i_\ell =k} x_{j_1,i_1}\otimes \cdots \otimes x_{j_\ell,i_\ell} \right),
\end{equation}
while the remaining components $x_{m,n}$ with $m\geq n$ are determined by
\begin{equation}\label{DSpropForm}
Y_n(x) = (I-\Lambda_n)^{-1} \, \Lambda_n V^{(n)}(x), 
\end{equation}
where $Y_n(x)^t=(x_{n,1}, \ldots, x_{n,n})$ and $V^{(n)}(x)^t=(V^{(n)}(x)_j)_{j=1,\ldots,n}$
$$ V^{(n)}(x)_j = \sum_{k=2}^n \sum_{r_1+\cdots+r_k=n \atop s_1+\cdots+s_k=j} x_{r_1,s_1}\otimes\cdots \otimes x_{r_k,s_k}. $$
\end{thm}

\proof By \eqref{betaPXprop} we see that we have $x_{1,1}= 1+ a_1 \beta_{1,1}\circ x_{1,1}$,
which can be solved for $x_{1,1}$ provided the invertibility of $1-\Lambda_1=1- a_1 \beta_{1,1}$ holds.
The components $x_{m,n}$ with $m<n$ are determined by \eqref{xmlessn}, where the
right hand side only involves components $x_{r,s}$ with $r<m$ and $s<n$.
The equations for the remaining entries $x_{m,n}$ with $n\leq m$
break into systems of the form
$$ (I - \Lambda_m) \left( \begin{array}{c} x_{m,1} \\ \vdots \\ x_{m,m} \end{array}\right) =
\sum_j (\Lambda_m)_{ij} \left( \sum_{k=2}^m \sum_{r_1+\cdots+r_k=m \atop s_1+\cdots+s_k=j} x_{r_1,s_1}\otimes\cdots \otimes x_{r_k,s_k} \right)_j. $$
Provided that the invertibility condition for the transformation $I - \Lambda_m$ holds, these
systems provide unique values for all the components $x_{m,n}$ with $n\leq m$, as a function
of the components $x_{r,s}$ with $r<m$, that we have already been determined. 
\endproof

Notice that one does not expect the invertibility of $I - \Lambda_n(a,\beta)$ to hold
in general, as $\Lambda_n(a,\beta)$ involves compositions with the elements $\beta_{j,i}$.
The simplest example in which one has invertibility can be obtained by taking $\beta_{m,n}=0$
for $m\neq n$ and $I-\beta_{n,n}$ an invertible transformation in $\cP(n,n)$.

\medskip

As in the operad case, we can define $\cP_{\beta,P}(m,n)$ to be the $\K$-vector
spaced spanned by all the properad compositions
$$ x_{k,n} \circ (x_{j_1,i_1}\otimes \cdots \otimes x_{j_\ell,i_\ell} ) $$
with $j_1+\cdots+j_\ell =m$ and $i_1+\cdots i_\ell =k$ and where all 
the $x_{m,n}$ are components of the solution to the properad Dyson--Schwinger
equation determined by the choice of  $P(t)$ and of $\beta=(\beta_{m,n})$, with
the invertibility condition of Theorem \ref{thmsolnDSprop}. The $\cP_{\beta,P}(m,n)$
form a sub-properad of $\cP_{{\rm flow},\cV'}$ with the induced composition.

\medskip

\section{Renormalization of the halting problem}\label{haltSec}

In \cite{Man2}, Manin adapted the formalism of perturbative renormalization to
the halting problem in the theory of computation. To summarize Manin's approach,
the basic idea is that one considers non-computable functions as an analog of
divergent Feynman integrals. Renormalization consists of a  procedure of extraction
of finite values from divergences and in this setting Manin proposes that a
suitable form of renormalization should lead to a procedure of extraction of a
``computable part" from a non-computable function. 

\smallskip

While this whole program is, at present, still in its early stages and some
of the ideas involved are still speculative, we attempt here to give a 
sense of the general picture and formulate some of the main questions,
as we see them. We will also discuss how Dyson--Schwinger equations
may be accommodated in this general circle of ideas. While most of what
is discussed in this section is more speculative than the previous parts
of the paper, we regard this viewpoint as an important motivation for
further developments.

\medskip

\subsection{The halting problem: regularization and renormalization}

We describe here a setting for renormalization of the halting problem
which is a minor elaboration on the procedure described in \S 3 of \cite{Man2}.
We do not add any significant new development beyond what is outlined as
a possible program in Manin's work \cite{Man2}, but we will try to identify 
some specific questions that we hope to return to in future work. 

\medskip

\subsubsection{Partial recursive functions and Hopf algebras}

We first pass from the class of primitive recursive functions considered above 
to the larger class of partial recursive functions. These have a similar presentation
as the one recalled in \S \ref{PrimRecSec} above, with the same basic functions
(successor, constants, and projections) and, in addition to the three elementary
operations $\c,\b,\r$ of composition, bracketing and recursion, an additional $\mu$
operation that assigns to an input function  $f: \N^{n+1} \to \N$ assigns an output
$$ h: \N^n \to \N, \ \ \  h  (x_1,\ldots,x_n) = \min \{ x_{n+1}\,|\, f(x_1,\ldots,x_{n+1})=1 \}, $$
with domain 
$$ \cD(h)=\{ (x_1,\ldots,x_n)\,|\, \exists x_{n+1}\geq 1\,:\,
f(x_1,\ldots,x_{n+1})=1, \, \text{ with } \, (x_1,\ldots,x_n,k)\in \cD(f), \forall k\leq x_{n+1} \}. $$
According to Church's thesis the partial recursive functions obtained in this way
are exactly the semi-computable functions, namely those for which there exists a 
program that, for all $x\in \cD(f)$, computes $f(x)$ and which either computes zero
or runs for an infinite time when $x\notin \cD(f)$, see \S V of \cite{Man-logic}. Notice
that it is the presence of the additional operation $\mu$ that allows for the
construction, starting from the basic functions, of functions that are only partially
defined and semi-computable.

\smallskip

Correspondingly, we enrich the Hopf algebras of flow charts by additional
vertex decorations by the $\mu$ operations. %
As in the case of the other operations, this requires defining a version
of the $\mu$ operation for vertices of valence greater than two. For
example, a possible choice may be to use the higher-arity bracketing operation
and an application of $\mu$ coordinatewise. We shall
still use the notations $\cH^{nc}_{{\rm flow}, \cR'}$ and $\cH^{nc}_{{\rm flow}, \cV'}$,
for the resulting Hopf algebras with admissible decorations by partial recursive 
functions or with vertex decorations, respectively, and 
$\cH_{{\rm flow}, \cR'}$ and $\cH_{{\rm flow}, \cV'}$ for their commutative
quotients.

\medskip

\subsubsection{Partial recursive functions and the halting problem}

As in \cite{Man2}, we regard the halting problem as a particular instance of
the general problem of recognizing whether a given $k \in \N$ belongs to 
the domain of definition $\cD(f)$ of a partial recursive function $f$. The halting
problem corresponds to the case where, given a partial recursive function $f$
and a program $P$ computing it, one considers the partial recursive
function $t(P)$, the computation time, with domain $\cD(t(P))=\cD(f)$,
see \cite{Man2}, \S 3.1.(a) and \S 2.1, and also \S IX of \cite{Man-logic}.


\medskip

\subsubsection{Algebraic renormalization}
The basic algebraic formalism for perturbative renormalization, in its
formulation as given in \cite{EFGK} (see also \S 5 of \cite{Mar}), consists of
an {\em algebraic Feynman rule} 
\begin{equation}\label{phiFey}
\phi: \cH \to \cB
\end{equation}
which is a {\em morphism of commutative
algebras} from a {\em commutative Hopf algebra} $\cH$ to a {\em
Rota--Baxter algebra} $\cB$ of weight $-1$.

\smallskip

Recall that a commutative Rota--Baxter algebra of weight $\lambda$ is 
a commutative, associative algebra $\cB$ endowed with a linear operator
$T$ on it that satisfies the identity
\begin{equation}\label{RotaBaxter}
T(x)T(y)=T(xT(y)) +T(T(x)y) + \lambda \, T(xy).
\end{equation}
The field of convergent Laurent series in one variable with 
the projection $T$ onto the polar part is an example of Rota--Baxter
algebra of weight $\lambda=-1$, which is widely used in
renormalization theory, \cite{CK}, \cite{CoMa}.
The Rota--Baxter operator $T$ of weight $-1$ determines a splitting
of $\cB$ into two commutative unital subalgebras, $\cB_+=(1-T)\cB$
and $\cB_-$ given by $T\cB$ with a unit adjoined. The fact that $\cB_\pm$
are indeed algebras and not just vector spaces follows from the 
Rota--Baxter identity \eqref{RotaBaxter}.

\smallskip

The BPHZ renormalization of the algebraic Feynman rule \eqref{phiFey} is
then obtained by showing the existence of a multiplicative factorization 
$$ \phi = (\phi_- \circ S) \star \phi_+, \ \ \ \text{ with } \ \ \
\phi_\pm: \cH \to \cB_\pm, $$
obtained inductively via the BPHZ preparation formula (see \cite{CK}, \cite{EFGK})
\begin{equation}\label{BPHZ}
\phi_-(x) = -T (\phi(x) + \sum \phi_-(x') \phi(x'')), \ \ \  \text{ and } \ \  \
\phi_+(x) = (1-T) (\phi(x) + \sum \phi_-(x') \phi(x'')),
\end{equation}
where $x\in \cH$, with $\Delta(x)=1\otimes x+x\otimes 1 + \sum x' \otimes x''$. 
The factorization is unique if normalized by $\epsilon_-\circ \phi_-=\epsilon$,
where $\epsilon_-:\cB_-\to \C$ is the augmentation map and $\epsilon$ is
the counit of $\cH$.
Again, it is the Rota--Baxter property of $(\cB,T)$ that ensures that $\phi_\pm$
obtained in this way are algebra homomorphisms. The homomorphism
$\phi_+: \cH \to \cB_+$ is the {\em renormalized} algebraic Feynman rule,
while $\phi_: \cH \to \cB_-$ is the {\em counterterm}, the divergence.

\medskip

\subsubsection{An algebraic Feynman rule for the halting problem} %
We recall here a proposal made by Manin in \cite{Man2} of a possible
construction of an algebraic Feynman rule in the setting of the halting problem. %

\smallskip

We let $\cB$ be the algebra of functions $\Phi: \N^m \to \cM(D)$ from
$\N^m$, for some $k$, to the algebra $\cM(D)$ of analytic functions
in the unit disk $D=\{z\in \C\,:\, |z|<1 \}$. Since the functions $\Phi$
can have different domains $\N^k$, the sums and products are 
defined by extending them to functions $\Phi(k_1,k_2,\ldots,k_n,\ldots)$ 
from a common domain $\N^\infty$, which depend on only finitely
many variables, $\Phi=\Phi\circ \pi^\infty_m$, with $\pi^\infty_m :\N^\infty \to \N^m$.
The Rota--Baxter operator $T$ on $\cB$ is given by componentwise (for
each $\underline{k}$) projection onto the polar part at $z=1$. Thus,
$\cB_+$ consists of functions from $\N^\infty$ (depending only on finitely
many $k_i$) to $\cM(D)_+$, the algebra
of meromorphic functions in $D$ that extend continuously at $z=1$, while
$\cB_-$ consists of functions that are either constant or have a pole at $z=1$. 

\smallskip

We define algebraic Feynman rules for the Hopf algebras 
$\cH_{{\rm flow}, \cR'}$ and $\cH_{{\rm flow}, \cV'}$. In the
case of $\cH_{{\rm flow}, \cR'}$, where both vertices and flags in the
trees are labeled by admissible labelings, there is a well defined
partial recursive function $f$ associated to each labelled rooted tree $\tau$ 
in $\cH_{{\rm flow}, \cR'}$, namely the function $f$ labeling the unique 
outgoing flag of $\tau$. Let us first assume that the partial recursive
function $f$ has range in $\N$. We extend this $f:\N^m \to \N$ to a function
$\bar f: \N^m \to \Z_{\geq 0}$ that computes $f(x)$ at $x\in \cD(f)$ and 
takes value $0$ at $x\notin \cD(f)$. 

\smallskip

We assign to a tree $\tau$ that computes $f$ 
an element in $\Phi_\tau(\underline{k},z)$ in $\cB$, defined (as in \S 3 of \cite{Man2})  by
\begin{equation}\label{Phikfz}
\Phi_\tau(\underline{k},z)=\Phi(\underline{k},f,z):= 
\sum_{n\geq 0} \frac{z^n}{(1+ n \bar f(\underline{k}))^2}.
\end{equation}
The resulting function $\Phi_\tau(\underline{k},\cdot)\in \cM(D)$ has a pole at $z=1$
iff $\underline{k}\notin \cD(f)$. We extend this definition mutiplicatively to cases where the output of the
flow chart $\tau$ is a partial recursive function $f:\N^k \to \N^\ell$, by setting
$\Phi_\tau(\underline{k},z)= \prod_{j=1}^\ell  \Phi(\underline{k},f_j,z)$. In this way, the
resulting function has a pole at $z=1$ if at least one of the $f_j$ does, that is, iff
$\underline{k}\notin \cD(f_1)\cap \cdots \cap \cD(f_\ell) = \cD(f)$, consistently with the above.  
We then extend the definition multiplicatively to the case of a monomial
$\tau_1 \cdots \tau_n$ in $\cH_{{\rm flow}, \cR'}$ and additively to linear combinations of monomials.
Thus, we obtain an algebra homomorphism $\Phi: \cH_{{\rm flow}, \cR'} \to \cB$.

\smallskip

We write $\Phi(\underline{k},f,z):=\Phi_f(\underline{k})(z)$ for the function $\Phi_f(\underline{k})\in
\cM(D)$ of $z\in D$, associated to a given $\underline{k}\in \N^\infty$ and depending on 
the output $f$ of a given flow chart, with $f$ a partial recursive functions. 
If we denote by $\tau$ the decorated
planar rooted tree describing the flowchart, we equivalently write $\Phi(\underline{k},\tau,z)$
or $\Phi_{\underline{k},\tau}(z)$.

\smallskip

The specific form \eqref{Phikfz} of the proposed Feynman rule does not appear to
be so important, as much as the two properties:
\begin{itemize}
\item The function $\Phi(\underline{k},f,\cdot)\in \cM(D)$ has a pole at $z=1$
iff $\underline{k}\notin \cD(f)$
\item The function $\Phi(\underline{k},f,\cdot)$ depends on $\underline{k}$ through
the values $\bar f(\underline{k})$ and these values can be reconstructed from the
function $\Phi(\underline{k},f,z)$.
\end{itemize}
Any other choice of a function with these properties would be a suitable
Feynman rule according to \cite{Man2}. Thus, it would seem perhaps 
more convenient to consider the set of all such functions, possibly up
to a suitable equivalence relation. We will discuss more below why it
may be necessary to introduce an equivalence relation on Feynman rules.
This has an analog in the quantum field theory setting, where the classification
of divergences in terms of a Galois theory based on the Tannakian formalism
was obtained in \cite{CoMa2} precisely by introducing a suitable equivalence
relation on algebraic Feynman rules. We will not develop this approach further
in the present paper, but we hope to return to it elsewhere.

\smallskip

In Manin's idea a possible renormalization of the halting problem
should provide a way of extracting a computable part from the halting problem for partial
recursive functions that accounts for ``subdivergences" created by subroutines within
the flowcharts. Though, at present, how exactly this should be achieved is not yet fully 
understood, in the procedure described in \cite{Man2} 
one would proceed by applying the BPHZ factorization \eqref{BPHZ} 
to the chosen algebraic Feynman rule. The result of factorization consists
of two new algebra homomorphisms from the same Hopf algebra to
the two parts of the target Rota--Baxter algebra. The main question
is then the interpretation of both pieces in terms of computability and
partial recursive functions. 

\smallskip

When we apply the BPHZ procedure to $\Phi$ defined as in \eqref{Phikfz}, we obtain
\begin{equation}\label{Phicoprod}
 \Phi_-(\underline{k},f_\tau,z) = -T (\Phi(\underline{k},f_\tau,z)+ \sum_C \Phi_-(\underline{k},f_{\pi_C(\tau)},z) \Phi(\underline{k},f_{\rho_C(\tau)},z)). 
\end{equation} 
Because $f=f_\tau$ is the output function computed by the flow chart, that is, the label
of the outgoing flag of $\tau$, we have $f_{\rho_C(\tau)}=f_\tau$, since these trees have
the same root and outgoing flag with the same label. Thus, the expression above is simply
of the form
\begin{equation}\label{PhisumC}
 \Phi_-(\underline{k},f_\tau,z) = -T \left(\Phi(\underline{k},f_\tau,z) (1+\Phi_-(\underline{k},\sum_C f_{\pi_C(\tau)},z)) \right). 
\end{equation} 
This has the effect of considering not only the pole at $z=1$ that is caused by $f_\tau$ itself, but
also those that come from subdivergences created by all the partial recursive functions 
$f_{\pi_C(\tau)}$ computed in the intermediate steps of the computation performed by the flow
chart $\tau$, which are the outputs of the ``pruned parts" $\pi_C(\tau)$ of the admissible cuts on $\tau$.

\smallskip

This suggests that the main role of the negative piece of the Birkhoff
factorization should be understood as a way of detecting the presence 
of a pole not only in the function $\Phi(\underline{k},f,z)$ itself, but in 
the analogous functions associated to the results $f_{\pi_C(\tau)}$
and $f_{\rho_C(\tau)}$ computed by subroutines of the program $\tau$
that computes $f$.

\smallskip

The role of ``subdivergences" in programs computing partial recursive
functions is interesting, for example because one can have programs 
that compute total functions as output, but involve partial functions,
so that the graph itself may be divergence-free and still contain
subdivergences, as the next example (suggested to us by Joachim Kock)
shows.

\begin{ex}\label{subdivex}{\rm
The identity function can be computed by a tree with one $\c$ node
and one $\mu$ node, as a composition of a successor function followed 
by a partial predecessor function, $\mu(| y+1-x|)$, undefined at $0$ and
$x-1$ for $x>0$. 
}\end{ex}

Thus, replacing the polar part of $\Phi(\underline{k},f,z)$ with the 
polar part $\Phi_-(\underline{k},f_\tau,z)$ can be understood as a
way to take into account all possible (and possibly hidden) subdivergences.

\smallskip

It is more difficult to obtain a good interpretation of the positive
piece of the Birkhoff factorization. The analogy with quantum field
theory suggests that it is not the value itself of $\Phi_+(\underline{k},f_\tau,z)$
that matters. After all, even in the original physical setting, there is an
intrinsic ambiguity in the renormalized values, coming from the action
of the renormalization group. The main question of Manin's approach
here appears to be the following (loosely formulated).

\begin{ques}\label{frenques}{\rm Is there a new function $f_{\rm ren}$, 
which is {\rm primitive
recursive}, such that
\begin{equation}\label{freneq}
 \Phi_+(\underline{k},f_\tau,z) = \Phi(\underline{k},f_{\rm ren},z), 
\end{equation} 
namely, is $\Phi_+$ equal to the original $\Phi$ of another function,
this time primitive instead of partial recursive?
}\end{ques}

If one could answer this question positively, then one could think of
the new function $f_{\rm ren}$ as an ``extraction of a computable 
piece" from the partial recursive function $f$ (and the program $\tau$
that computes it). Upon close inspection, it is easy to see that, as
naively formulated in Question \ref{frenques}, it may not be possible
in general to obtain such an $f_{\rm ren}$. However, one can perhaps
formulate a better question of whether an identification \eqref{freneq}
holds after considering the algebraic Feynman rules up to an equivalence 
relation, as with the gauge transformations
on germs of meromorphic functions considered in the quantum field
theory setting in \cite{CoMa2}. We will not develop this issue further
in the present paper, but we leave it as an open question.

\subsubsection{The case of $\cH_{{\rm flow}, \cV'}$}

We make here a sketch of a possible suggestion for a 
variation on the construction above, adapted to define an algebraic Feynman rule
on $\cH_{{\rm flow}, \cV'}$, where the flags are not labelled and only vertices are.
In this case, in order to define something like \eqref{Phikfz}, we need a way of assigning 
inputs to the flow charts. A reasonable choice is to use the basic functions as input. We then
define the algebraic Feynman rule as 
\begin{equation}\label{Phisigma}
\Phi(\underline{k},\tau,z):= \prod_\sigma  \Phi(\underline{k},f_{\tau,\sigma},z),
\end{equation}
where the product is over $\sigma$ ranging over the set of all possible 
functions $\sigma: E_{ext}^{in}(\tau) \to \{ s, c, \pi \}$ that label the incoming
external edges of $\tau$ by basic functions (successor, constant, or projection). %
The target of $\sigma$ contains all the possible basic functions (of type $s$, $c$,
or $\pi$). Since this is an infinite set (there are infinitely many possible choices
of projection functions $\pi_i^n$), if one wants to have a finite product in
\eqref{Phisigma} one may have to restrict $\sigma$ by making a priori
choices of smaller sets of basic functions as target, or else introduce further conditions
on the class of functions $\Phi$ that would ensure convergence. %
The partial recursive function $f_{\tau,\sigma}$ is the output of the flow chart
given by the tree $\tau$ with inputs assigned by $\sigma$.
Each $\Phi(\underline{k},f_{\tau,\sigma},z)$ is computed as in  \eqref{Phikfz},
and \eqref{Phisigma} is extended to arbitrary elements of the Hopf algebra as
before.  In this setting, the function $\Phi_\tau(\underline{k},z)$ has a pole at $z=1$ iff
there is at least a choice of a basic input $\sigma$ for which $\underline{k}\notin \cD(f_{\tau,\sigma})$.

\smallskip

With this setting, the BPHZ formula for the algebraic Feynman rule \eqref{Phisigma}
becomes more interesting than in the case of \eqref{PhisumC}. We have again \eqref{Phicoprod},
in the form
\begin{equation}\label{divergencetau}
 \Phi_-(\underline{k},\tau,z) = -T (\Phi(\underline{k},\tau,z)+ \sum_C \Phi_-(\underline{k},\pi_C(\tau),z) \Phi(\underline{k},\rho_C(\tau),z)),
\end{equation} 
with $\Phi(\underline{k},\rho_C(\tau),z)$ and $\Phi(\underline{k},\pi_C(\tau),z)$ again
computed as in \eqref{Phisigma}, 
but this time it is no longer true that $f_\tau=f_{\rho_C(\tau)}$ because on the tree 
$\rho_C(\tau)$ we are using the new inputs given by basic functions and not the input
coming from the output of $\pi_C(\tau)$.  The divergence \eqref{divergencetau} 
here combines the divergences of the $f_{\tau,\sigma}$ with 
divergences coming from partial recursive functions $f_{\pi_C(\tau),\sigma}$ and
$f_{\rho_C(\tau),\sigma}$ that flowcharts $\pi_C(\tau)$ and $\rho_C(\tau)$ compute
starting from inputs of basic functions.

\medskip

\subsection{Dyson--Schwinger equations for the halting problem}

Finally, we tentatively propose a possible role for Dyson--Schwinger equations
in the context of Manin's approach to the halting problem. These remarks
are also, at this stage, very speculative, as they would depend for a more
precise interpretation upon a better understanding of the main questions
about the Birkhoff factorization discussed earlier in this section.

\smallskip

If we work with the Feynman rule described above on the Hopf algebra
$\cH_{{\rm flow}, \cV'}$ with only vertex decorations, then we can readily
consider Dyson--Schwinger equations as discussed in \S \ref{DSflowSec} above.
However, if we work with the definition of the algebraic Feynman 
rule $\Phi$ described in \eqref{Phikfz} on the Hopf algebra $\cH_{{\rm flow}, \cR'}$ 
of flow charts with flag decorations, we need to extend the grafting operators
appropriately  that are needed to define Dyson--Schwinger equations from 
$\cH_{{\rm flow}, \cV'}$ to $\cH_{{\rm flow}, \cV'}$. This can be done by assigning 
as output the empty function $f=\emptyset$, mapped to
the constant function $\Phi(\underline{k},\emptyset,z)\equiv 1$, as the output of any tree 
$\tau$ that is obtained through a grafting where the flag labels don't match. 

\smallskip

We can then consider a Dyson--Schwinger equation of the form \eqref{DSP}, with
$B^+=\sum_\delta B^+_\delta$, with $\delta\in \{ \b,\c,\r,\mu \}$ or a more general
system of Dyson--Schwinger equations \eqref{DSsystem}, with the condition that
the components $x_m$ of the unique solution span a Hopf subalgebra of 
$\cH_{{\rm flow}, \cR'}$. In such case, we
can restrict the algebraic Feynman rule $\Phi$ constructed above to this Hopf
subalgebra and still perform the BPHZ renormalization. 

\smallskip

One can then reformulate the same questions that we have briefly
discussed regarding Manin's approach to the halting problem, of how
to interpret the resulting negative and positive parts of the Birkhoff
factorization. In this case, the fact that one is restricting attention to
the Hopf subalgebra generated by the components of the solution to
the Dyson--Schwinger equation means that  the negative part
$\Phi_-(\underline{k},f_{x_m},z)$ (or $\Phi_-(\underline{k},f_{x_m,\sigma},z)$
in the case of $\cH_{{\rm flow}, \cV'}$) will now account for subdivergences
that belong to the same Hopf subalgebra determined by the Dyson--Schwinger 
equation. This can be thought of, heuristically, as measuring the ``amount of 
non-computability" that can be produced by sub-flow-charts that are characterized
by a certain self-similarity property (defined by the Dyson--Schwinger
equation). An analog in this setting of Question \ref{frenques} on the
interpretation of the positive part of the Birkhoff factorization can also
be formulated in the Hopf subalgebra determined by the Dyson--Schwinger
equation.

\bigskip
\bigskip

\noindent {\bf Acknowledgment.} The first author was supported for this project by the 
Summer Undergraduate Research Fellowship (SURF) program of Caltech, 
through a Herbert J. Ryser fellowship. The second author is partially supported 
by NSF grants DMS-0901221, DMS-1007207, DMS-1201512, and PHY-1205440.
The second author acknowledges MSRI for hospitality and support. The authors are
especially grateful to Joachim Kock for many helpful comments and 
suggestions that significantly improved the paper.

\end{document}